\documentclass{aa} 
\usepackage{natbib}
\bibpunct{(}{)}{;}{a}{}{,}
\usepackage{graphicx}
\usepackage{float}
\usepackage{multirow}

\usepackage{txfonts}
\usepackage[normalem]{ulem}

\usepackage[dvipsnames]{xcolor}
\usepackage{soul}
\sethlcolor{orange}

\usepackage{hyperref} \hypersetup{ colorlinks=true, citecolor={blue} ,linkcolor=blue, filecolor=blue, urlcolor=blue, pdftitle={Overleaf Example}}

\begin{document} 

\newcommand{\mum}{$\upmu\rm{m}$}
\newcommand{\ms}{m\,s$^{-1}$}
\newcommand{\kms}{km\,s$^{-1}$}
\newcommand{\vsini}{$v\sin i$}
\newcommand{\vsinia}{$v\sin i_{\rm A}$}
\newcommand{\vsinib}{$v\sin i_{\rm B}$}
\newcommand{\vrad}{$v_{\rm rad}$}
\newcommand{\vmic}{$v_{\rm mic}$}
\newcommand{\vmac}{$v_{\rm th}$}
\newcommand{\msun}{M$_{\odot}$}
\newcommand{\msunpyr}{M$_{\odot}$\,yr$^{-1}$}
\newcommand{\rsun}{R$_{\odot}$}
\newcommand{\bl}{$B_{\ell}$}
\newcommand{\bd}{$B_{\rm D}$}
\newcommand{\ddeg}{$^{\circ}$}
\newcommand{\te}{$T_{\rm eff}$}
\newcommand{\logg}{$\log{g}$}
\newcommand{\rr}{$R_A/R_B$}
\newcommand{\lr}{$L_A/L_B$}
\newcommand{\prot}{$P_{\rm rot}$}

   \title{Star-disk interactions in the strongly accreting T Tauri Star S~CrA~N\thanks{PI (Perraut) Program 18AF12. Based on observations obtained at the Canada-France-Hawaii Telescope (CFHT) which is operated by the National Research Council of Canada, the Institut National des Sciences de l'Univers of the Centre National de la Recherche Scientifique of France, and the University of Hawaii.}}
   \author{H.~Nowacki           \inst{1}\fnmsep\thanks{\email{hugo.nowacki@univ-grenoble-alpes.fr}}
          \and E.~Alecian       \inst{1}
          \and K.~Perraut       \inst{1}
          \and B.~Zaire         \inst{2}
          \and C.P.~Folsom      \inst{3}
          \and K.~Pouilly       \inst{4}
          \and J.~Bouvier       \inst{1}
          \and R.~Manick        \inst{1}
          \and G.~Pantolmos     \inst{1}
          \and A.P.~Sousa       \inst{1}
          \and C.~Dougados      \inst{1}
          \and G.A.J~Hussain    \inst{5}
          \and S.H.P.~Alencar   \inst{2}
          \and J.B.~Le~Bouquin  \inst{1}
          }

   \institute{Univ. Grenoble Alpes, CNRS, IPAG, 38000 Grenoble, France
         \and Departamento de f\'isica, Universidade Federal de Minas Gerais, Belo Horizonte, MG, 31270-901, Brazil
         \and University of Tartu, Faculty of Science and Technology, Tartu Observatory, Estonia
         \and Department of Physics and Astronomy, Uppsala University, Box 516, SE-75120 Uppsala, Sweden
         \and Science Division, Directorate of Science, European Space Research and Technology Centre (ESA/ESTEC), Keplerlaan 1, NL-2201AZ Noordwijk, the Netherlands
         }

   \date{Received June 9$^\text{th}$, 2023; accepted August 3$^\text{rd}$, 2023}

  \abstract
   {Classical T Tauri Stars are thought to accrete material from their surrounding protoplanetary disks through funnel flows along their magnetic field lines. Among them, those with high accretion rates ($\sim 10^{-7}$\msunpyr) are ideal targets to test this magnetospheric accretion scenario in a sustained regime.
   }
   {We aimed at constraining the accretion-ejection phenomena around the strongly-accreting Northern component of the S~CrA young binary system (S~CrA~N) by deriving its magnetic field topology and its magnetospheric properties, and by detecting ejection signatures, if any.}
   {We led a two-week observing campaign on S~CrA~N with the ESPaDOnS optical spectropolarimeter at the Canada-France-Hawaii Telescope. We recorded 12 Stokes $I$ and $V$ spectra over 14 nights. We computed the corresponding Least-Square Deconvolution (LSD) profiles of the photospheric lines and performed Zeeman-Doppler Imaging (ZDI). We analysed the kinematics of noticeable emission lines, namely He~I $\lambda 5876$ and the four first lines of the Balmer series, known to trace the accretion process.}
   { We found that S~CrA~N is a low-mass (0.8~\msun), young ($\sim$~1~Myr), and fully convective object exhibiting a strong and variable veiling (with a mean value of 7~$\pm$~2), which suggests that the star is in a strong accretion regime. These findings could indicate a stellar evolutionary stage between Class I and Class II for S~CrA~N. We reconstructed an axisymmetric large-scale magnetic field ($\sim$~70\% of the total energy), primarily located in the dipolar component but with significant higher poloidal orders.
   From the He~I $\lambda 5876$ narrow emission component radial velocity curve, we derived a stellar rotation period of $P_* = 7.3 \pm 0.2$ days. We found a magnetic truncation radius of $\sim$ 2~R$_*$ which is significantly closer to the star than the corotation radius of $\sim$~6~R$_*$, suggesting that S~CrA~N is in an unstable accretion regime. The truncation radius being quite smaller than the size of the Br$\gamma$ line emitting region, as measured with the GRAVITY interferometer ($\sim$ 8~R$_*$), supports the presence of outflows, which is nicely corroborated by the line profiles presented in this work. }
   {The findings from spectropolarimetry are complementary to those provided by optical long-baseline interferometry, allowing us to construct a coherent view of the innermost regions of a young, strongly accreting star. Yet, the strong and complex magnetic field reconstructed for S~CrA~N is inconsistent with the observed magnetic signatures of the emission lines associated to the post-shock region. We recommend a multi-technique, synchronized campaign of several days to put more constrains on a system that varies on a $\sim$ 1 day timescale.}

   \keywords{Stars: variables: T Tauri --
                Stars: individual: S~CrA~N --
                Stars: magnetic field --
                Techniques: spectroscopic, polarimetric --
                Accretion, accretion disk
               }

   \maketitle

\nolinenumbers
\section{Introduction}
Protoplanetary disks are gas- and dust-rich regions, mostly constituted of hydrogen and helium gas, with a small fraction of their content in dust grains. As the reservoir from which matter is accreted onto the star and planets are built, they set the initial conditions for planet formation. In the innermost regions of these disks, accretion flows, winds, and outflows are essential to control angular momentum, alter the gas content, and drive the dynamics of the gas \citep{alexander_dispersal_2014}. While they are key elements for understanding locally and globally the disk dynamics and its evolution, the physical structure, the exact nature of the processes acting in these inner regions, as well as their interplay, are still poorly constrained and not well understood.

Classical T Tauri stars (CTTS) are young suns (< 2 M$_\odot$) that still accrete from their surrounding disk. The most accepted scenario is that accretion is funneled by the stellar magnetic field (\citealp{hartmann_accretion_2016}; \citealp{bouvier_magnetospheric_2007}) that is strong enough to truncate the circumstellar disk at a few stellar radii from the central star. This magnetic field drives the ionized gas of the inner disk that falls onto the central object, and forms shocks and hot spots at high latitudes on the stellar surface \citep[]{camenzind_magnetized_1990, koenigl_disk_1991, espaillat_multiwavelength_2022}. These phenomena shape many properties of the accreting objects and can be investigated by spectroscopy, spectropolarimetry, and photometry as their host stars exhibit excess continuum emission in the optical and near-infrared ranges, hot and cold spots, as well as broad, intense, and variable emission lines in the visible and near-infrared ranges (e.g., \citealp[]{alencar_accretion_2012, alencar_inner_2018, sousa_star-disk_2021, sousa_new_2023}).

The accretion regime depends on the large-scale magnetic field strength (i.e., usually a dipole), the angle between the dipolar magnetic-field and stellar rotation axes, and the mass accretion rate (\citealp{kulkarni_accretion_2008}; \citealp{blinova_boundary_2016}). It can be either stable when occurring through two funnels, one per hemisphere, or unstable when several equatorial tongues penetrate the stellar magnetosphere. These tongues are transient on timescales of the stellar rotation period and can coexist with stable accretion funnels \citep{kulkarni_accretion_2008, pantolmos_stable_2022}. The transition between these regimes strongly depends on the mass accretion rate: unstable accretion is expected to be observed mostly in strong accretors rather than in low accretors \citep{blinova_boundary_2016}. Strong accretors allow us to probe a different accretion regime, through which all low-mass stars should go during their early pre-main sequence (PMS) evolution as they move away from the protostellar phase \citep{baraffe_self-consistent_2017}. Yet, they  have been poorly explored until now, because their variability in the optical domain complicates their spectral analysis.

Until recently, the structure of the magnetosphere has been mostly probed through indirect observations thanks to the measurements of magnetic field strength and topology \citep{donati_spectropolarimetric_1997}, and mass accretion rate estimates (\citealp{manara_penellope_2021}; \citealp{alcala_giarps_2021}). The drastic improvement of sensitivity of optical long-baseline interferometers has opened a new promising way to probe the interaction between the young stars and their inner disks. The K-band interferometric beam-combiner GRAVITY at the VLTI \citep{gravity_collaboration_first_2017} makes it possible to spatially resolve the Br\,$\gamma$ line emitting region for a few T Tauri stars \citep{gravity_collaboration_gravity_2023}. Combined with spectropolarimetry, this appears very promising as it allows comparing the size of the Br\,$\gamma$ line emitting region with that of the magnetosphere derived by spectropolarimetry, and thus to investigate the accretion-ejection processes at (sub-)astronomical unit scale (\citealp{bouvier_probing_2020}; \citealp{gravity_collaboration_measure_2020}, \citeyear{gravity_collaboration_gravity_2023}).

With the aim of studying the peculiar accretion regime of a strong accretor through complementary observing techniques, we focus our work on the North component of the young binary system S Coronae Australis (S~CrA~N) as it is one of the strongest accretors of the GRAVITY T Tauri sample ($\Dot{M}\sim$~10$^{-7}$~\msunpyr; \citealp{gahm_s_2018} ; \citealp{sullivan_s_2019}). To complete the GRAVITY data set and better understand the accretion-ejection phenomena at play, we have conducted an observing campaign with the optical spectropolarimeter ESPaDOnS. 

S~CrA is a T~Tauri binary system whose components are coeval (\citealp{gahm_s_2018} and references therein) and separated by about 1.4” \citep{reipurth_visual_1993}. Due to the inverse P Cygni profiles of its hydrogen H$\gamma$ and H$\delta$ lines, \cite{rydgren_observations_1977} classified this system as a YY~Ori object, a sub-class of PMS stars characterized by their excess in the Ultra-Violet (UV) and the inverse P Cygni structure in the high orders of the Balmer series \citep{walker_studies_1972}. Many spectroscopic studies in the optical and near-infrared ranges of S~CrA have been reported in the literature and focused on characterizing its veiling \citep{prato_astrophysics_2003}, its variability (\citealp{edwards_line_1979}; \citealp{sullivan_s_2019}), and more generally on the accretion-ejection processes at play \citep{gahm_s_2018}. From these spectral analyses, S~CrA~N appears to be highly obscured by a variable veiling and to exhibit a high mass accretion rate onto the star, suggesting that this star might be transitioning between the Class I and Class II stages of stellar evolution.

S~CrA has been part of the ALMA survey of protoplanetary disks in the Corona Australis region \citep{cazzoletti_alma_2019}; a continuum emission at 1.3~mm with a half-width of half-maximum of about 0.22" is detected around S~CrA~N; it exhibits neither clear substructures nor an inner cavity with radii larger than 25~au. S~CrA~N has also been observed in the mid-infrared range with the MIDI instrument of the VLTI (\citealp{schegerer_tracing_2009}; \citealp{varga_vltimidi_2018}): a half-flux radius of the continuum emitting region of about 1.4~au was derived, but the observations did not allow to accurately constrain the disk properties. More recently, the near-infrared continuum emission of S~CrA~N has been partially resolved in the H-band with PIONIER \citep{anthonioz_vltipionier_2015} and in the K-band with GRAVITY \citep{gravity_collaboration_wind_2017, gravity_collaboration_gravity_2021}, and a half-flux radius of about 0.1~au has been derived for this continuum emission. Moreover, by fitting the continuum K-band interferometric
data obtained with GRAVITY, \cite{gravity_collaboration_gravity_2021} derived an inclination of the inner disk of 27~$^\circ \pm$~3$^\circ$.
 Thanks to the spectrometric capabilities of GRAVITY, the Br$\gamma$ emitting region close to the star has also been partially resolved (0.06-0.07~au), appearing more compact than the continuum (\citealp{gravity_collaboration_wind_2017}; \citealp{gravity_collaboration_gravity_2023}).

In this paper, we report on the spectropolarimetric campaign we led on S~CrA~N in the optical range to complete the GRAVITY near-infrared interferometric study of \cite{gravity_collaboration_gravity_2023}. The observations and the data processing are described in Section~\ref{Sect:obs}. We present our results in Section~\ref{Sec:Results} and discuss them in Section~\ref{Sec:discuss}.

\section{Data}
\label{Sect:obs}

\subsection{Observations}
\begin{table}[t]
\caption{Log of the ESPaDOnS observations of S~CrA~N including the Date of observation, Heliocentric Julian Date (HJD), total exposure time ($t_{\rm exp}$), Signal to Noise Ratios (SNR) as the average value computed in the order centered on 581 nm, and seeing as measured by the Mauna Kea Atmospheric Monitor (MKAM) at the time of observation.}
\label{table:log}   
\centering                    
\begin{tabular}{c c c c c c c}      
\hline\hline                
Date & HJD & $t_{\rm exp}$ & SNR(I) & SNR(V) & Seeing \\ 
(2018) & (-2458000) & (s) & & &(arcsec)\\
\hline                       
  June 21 & 290.96670 & 3562 & 159 & 121 & 0.43 \\     
  June 22 & 291.97893 & 3600 & 90 & 68 & 0.81 \\
  June 23 & 292.95395 & 3600 & 118 & 93 & 0.58 \\
  June 24 & 293.99200 & 3600 & 117 & 88 & 0.63 \\
  June 25 & 294.93505 & 3600 & 124 & 94 & 0.52 \\
  June 26 & 295.97417 & 3448 & 156 & 116 & 1.30 \\
  June 27 & 296.94506 & 3600 & 114 & 84 & 1.35 \\
  June 28 & 297.91663 & 3600 & 121 & 88 & 1.05 \\
  June 29 & 298.94801 & 3600 & 115 & 86 & 0.50 \\
  June 30 & 299.95373 & 3600 & 117 & 86 & 0.60 \\
  July 01 & 300.90393 & 3600 & 114 & 80 & 0.62 \\
  July 04 & 303.93473 & 3600 & 100 & 67 & 0.57 \\
\hline                                   
\end{tabular}
\end{table}

S~CrA~N was observed with ESPaDOnS \citep{donati_espadons_2003}, the spectropolarimeter at Canada France Hawaii Telescope (CFHT), for 11 consecutive nights from the 21$^\text{st}$ of June to the 2$^\text{nd}$ of July 2018, plus one last observation on the 4$^\text{th}$ of July. Altogether, they represent a set of 12 observations over a spectral range from 367~nm to 1048~nm, with a spectral resolution $R\sim 65\,000$. The log of the observations is given in Table~\ref{table:log}.
Each observation consists of Stokes $I$, $V$, and null ($N$) spectra per Heliocentric Julian Date (HJD). The Stokes $I$ parameter represents the total intensity of the light and allows for a classical spectroscopic analysis at a very high resolution. The Stokes $V$ parameter is the difference between left and right circularly polarized light, which provides the observer with information on the line-of-sight component of the magnetic field in the region where the light comes from. The $N$ signal is computed in a way that cancels the polarization of the incoming light. It is used to check for spurious polarization signals in the Stokes $V$ data: any detection in Stokes $V$ tallying with a $N$ signal significantly above its Root Mean Square (RMS) noise should be considered spurious. 

Since the separation between S~CrA~N and S~Cra~S is 1.4" while the average seeing at CFHT was 750 mas during our observations, we made sure that the contribution of S~CrA~S to the total observed flux was negligible (see Appendix~\ref{app:SCrAS} for the complete treatment). In the end, the average contribution of S~CrA~S is as low as 1.7\%, with 3 observations exceeding a contribution of 5\% : June 26, 27 and 28 with 9.0\%, 9.3\% and 5.3\%, respectively. We considered these contributions negligible for the rest of the study but stayed vigilant to any unexpected behaviour that would come out from any of these three observations.

\subsection{Data reduction}
The data were reduced automatically through the Libre ESpRIT procedure (see \citealp{donati_spectropolarimetric_1997}). This reduction also includes the normalization of the Stokes $I$ continuum, which was imperfect in the case of S~CrA~N due to the high activity of the object. The numerous intense emission lines led to a biased estimation of the continuum level in some spectral windows. We used the \texttt{SpeNT} code developed by \cite{martin_first_2018} to refine the normalization of our spectra until the continuum level varied by no more than the RMS signal outside any line in Stokes $I$. The code fits a third order spline to the continuum considering a $\sigma$-clipping procedure to automatically reject absorption/emission lines. Nevertheless, this procedure can be manually adjusted by visually rejecting any fixed point of the spline should the considered portion of the spectrum be highly variable. For S~CrA~N, we performed the fit on the average spectrum of all the Stokes $I$ spectra in our sample. Then, the resulting normalization was applied to all the spectra. This procedure proved to be essential to obtain good normalized spectra for S~CrA~N. 

\begin{figure}
    \centering
    \includegraphics[trim=20 5 0 10,clip,width=\linewidth]{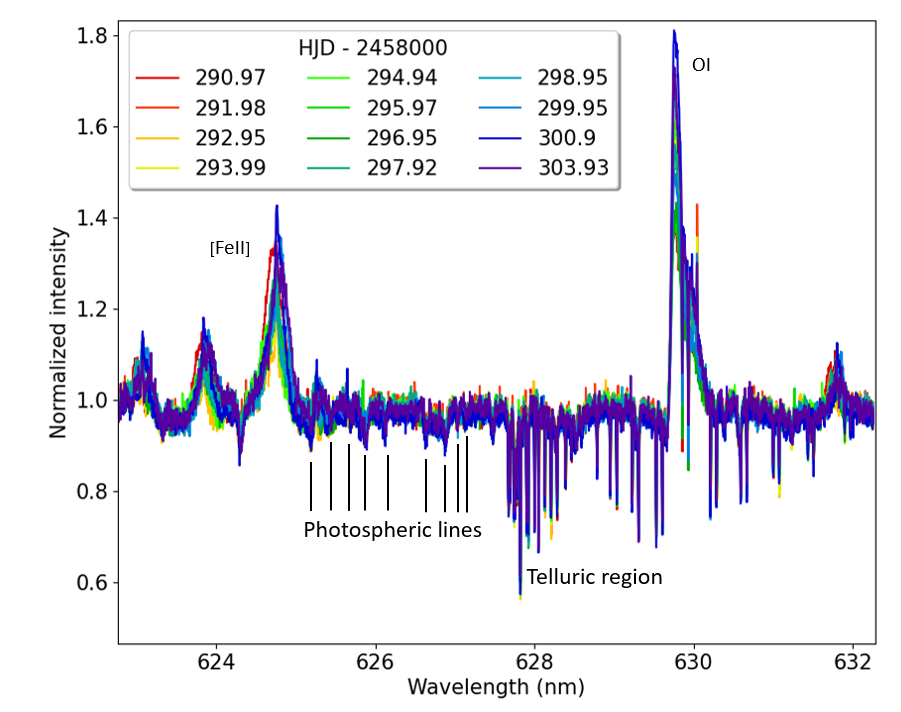}
    \caption{Portion of all the normalised spectra. Each color stands for one observation. The same color code will be used for the rest of the study.}
    \label{Fig:spectra_zoom}
\end{figure}

We recovered 12 good-quality spectra (one per night) displaying the usual telluric lines, photospheric lines, and classical emission lines for CTTS, as previously reported for this object \citep{gahm_s_2018}. We present a portion of the spectra in Fig.~\ref{Fig:spectra_zoom} where all these different features are visible at the same time. Photospheric lines are present but appear very weak on average (about 5\% of the continuum level) due to veiling over the whole range of observation. The emission lines are particularly intense on average, with variable shapes (e.g. the iron lines on the left of Fig.~\ref{Fig:spectra_zoom}). Many of the observed emission lines sensitive to the magnetic field exhibit significant Stokes $V$ counterparts, suggesting a strong magnetic field (these lines are presented in Appendix~\ref{app:spectra}). Finally, the telluric lines were not removed since no line of interest was located inside a telluric region.

\subsection{Least Square Deconvolution}
\label{Sec:Data}

One convenient way to look at these data is to compute their Least Square Deconvolution (LSD) profiles in both Stokes $I$ and Stokes $V$ spectra \citep{donati_spectropolarimetric_1997}: the photospheric absorption lines and their Stokes $V$ counterpart are weighted by their central wavelength, depth and Landé factor; then these weighted lines are averaged altogether over one observation. We built a mask that identifies the lines to include in this computation using the VALD database\footnote{VALD database : http://vald.astro.uu.se/}\citep{piskunov_vald_1995,ryabchikova_major_2015}: by scanning the average spectrum over-plotted to a synthetic spectrum produced with the MARCS stellar atmosphere model \citep{gustafsson_grid_2008}, we searched for undisturbed photospheric lines to be included. We paid particular attention not to include absorption lines contaminated by emission or telluric lines. Usually, the LSD $I$ profiles obtained exhibit distortions that are attributable to spots located at the photosphere level. In the case of LSD $V$, these profiles are shaped by the magnetic field along the line of sight at the surface of the star. The weak field approximation assumes that the broadening of the photospheric profiles due to the Zeeman effect (proportional to the local magnetic intensity along with the Landé factor and the central wavelength of a line) is negligible compared to all the other sources of broadening (instrumental broadening, thermal Doppler broadening, non-thermal Doppler broadening, microturbulent and macroturbulent velocities dispersions). Combined in quadratic sum, these effects correspond to a broadening of $\Delta v_\text{tot}~=~16.58$ km/s. For the specific case of S~CrA~N observed with ESPaDOnS, this $\Delta v_\text{tot}$ translates into $B\ll 10~\text{kG}$. In other words, the use of the LSD profiles is justified under the condition that the surface averaged magnetic field does not reach 10~kG or more. 

\begin{figure}
    \centering
    \includegraphics[trim=43 40 57 64, clip, width=\linewidth]{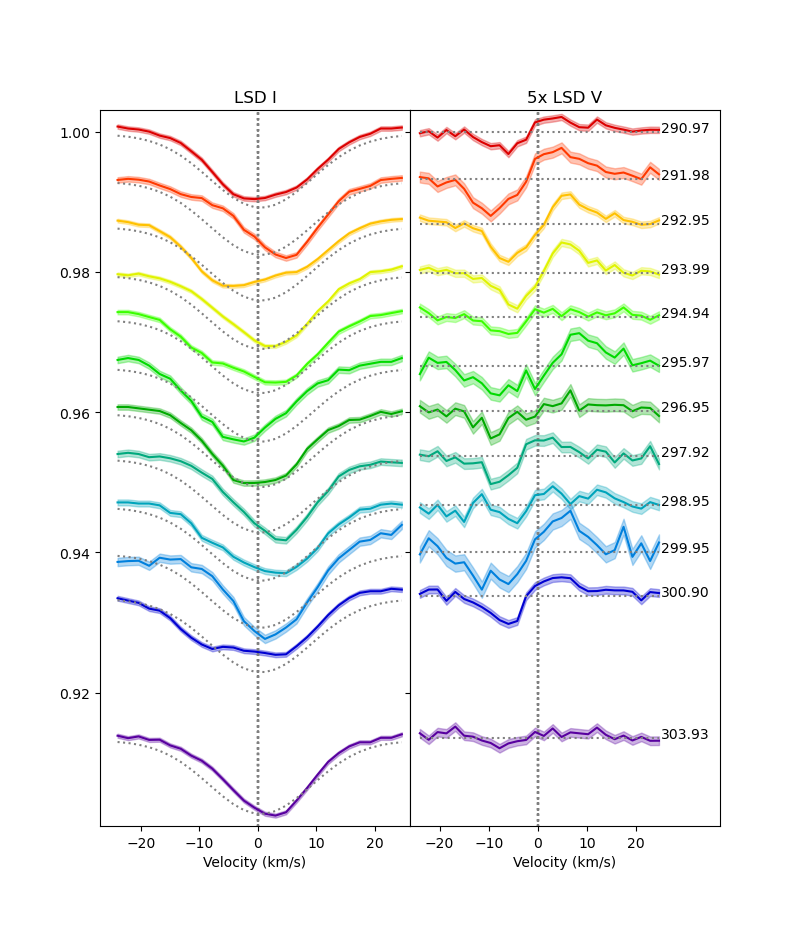}
    \caption{ LSD $I$ and $V$ profiles of S~CrA~N sorted by HJD (HJD - 2458000 mentioned on the right of each plot) normalized by the maximum EW, observed at date 303.93. The color coding is the same as in Fig.~\ref{Fig:spectra_zoom}. The spacing between two profiles stands for the time span between the observations. Each profile is represented with a solid colored line and is surrounded by its uncertainties in faded color. Black dotted lines represent the reference Voigt profile for the $I$ profiles and the continuum for the $V$ profiles. Dashed vertical lines represent the radial velocity of the star. $V$ profiles are magnified by a factor 5 for clarity.}
    \label{Fig:LSD_profiles}
\end{figure}

After applying the LSD computation with a Landé factor of 1.2 and a central wavelength of 500 nm to the complete Stokes $I$ and $V$ spectra, we obtain 12 set of LSD $I$, LSD $V$ and LSD $N$ profiles (Fig.~\ref{Fig:LSD_profiles}): one profile represents one observation date, which will be mentioned as (HJD - 2458000) hereafter (see Table \ref{table:log}), for convenience. We report that the LSD $N$ profiles show no spurious signal at any velocity for any observation.\\

\section{Results}
\label{Sec:Results}

\begin{table*}[t]
\caption{Stellar parameters of S CrA N.
}    
\label{table:StParams}      
\centering          
\begin{tabular}{l c c l l}    
\hline\hline
  Parameter & This work  & Literature & Method in the literature & Our method (if different) \\   
\hline
    \multirow{2}{*}{$T_\text{eff}~[K]$} & \multirow{2}{*}{$4300 \pm 100$ } & 4250 & Fit of photospheric lines\tablefootmark{a} & \multirow{2}{*}{Fit of photospheric lines} \\
     & & $4800 \pm 400$ & Comparison to spectral type standard star\tablefootmark{b}& \\[1ex]
    $v\sin{i}~[km/s]$ & $10 \pm 2$  & 12 & Fit of photospheric lines\tablefootmark{a} & \\[1ex]
    $r$ & $[4-11]$ & 8.3 & Fit of photospheric lines\tablefootmark{a} & \\[1ex]
    $v_r~[km/s]$ & $0 \pm 1$ & $0.9 \pm 2.5$ & Fit of photospheric lines\tablefootmark{a} & \\[1ex]
    \multirow{2}{*}{$d~[\text{pc}]$} & \multirow{2}{*}{$152.4 \pm 0.4$} & 129 & $uvby, \beta$ photometry\tablefootmark{c} & \multirow{2}{*}{"On-cloud" sub-region distance\tablefootmark{e}} \\
     & & $138 \pm 16$ & Light echoes analysis\tablefootmark{d} & \\[1ex]
    $L_*~[L_\sun]$ & $1.67 \pm 0.8$ & $2.29^{+0.76}_{-0.65}$ & Fit of SED\tablefootmark{b} & Distance-corrected value\tablefootmark{b}. \\[1ex]
    $M_*~[M_\sun]$ & $0.8 \pm 0.1$ & 0.7\tablefootmark{f} & From $T_\text{eff}$ assuming an age of 2 Myr & Position in HR diagram \\[1ex]
    $R_*~[R_\sun]$ & $2.3 \pm 0.6$ &  &  & Position in HR diagram \\[1ex]
    $P_*$~[days] & $7.3 \pm 0.2$ & $4.2 \pm 1$ & Periodicity in the He~I narrow component\tablefootmark{a} & \\ [1ex]
    $\log g~[\text{dex}]$ & $3.6 \pm 0.2$  &  &  & Direct computation \\[1ex]
\hline
\end{tabular}
\tablefoot{ References for the values and method in the literature.
\tablefoottext{a}{\cite{gahm_s_2018};}
\tablefoottext{b}{\cite{prato_astrophysics_2003};}
\tablefoottext{c}{\cite{marraco_distance_1981};}
\tablefoottext{d}{\cite{ortiz_observation_2010};}
\tablefoottext{e}{\cite{galli_corona-australis_2020};}
\tablefoottext{f}{\cite{sullivan_s_2019}}
}
\end{table*}

In this section, we present the methods used for the analysis of the set of data introduced above, and present our results.

\subsection{Fit of the Stokes $I$ spectra}
\label{SubSec:StellarParams}

\begin{figure}[ht]
    \centering
    \includegraphics[trim=42 14 65 54, clip, width=0.95\linewidth]{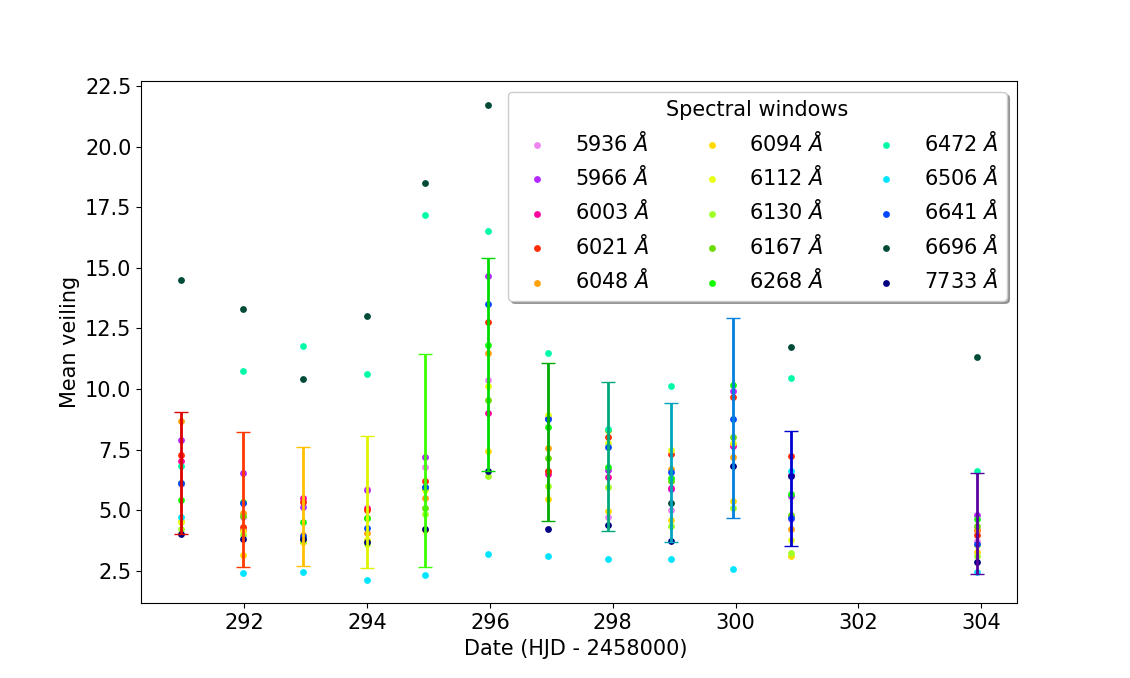}
    \caption{Evolution of the average optical veiling with time. For each date, each color point stands for the mean veiling over one spectral window whose center value is indicated in the caption. The error bars are centered on the average of the values computed over the 15 spectral windows and spread over their standard deviation.}
    \label{Fig:veiling}
\end{figure}

We used the \texttt{ZEEMAN} code \citep[]{landstreet_magnetic_1988, wade_lte_2001, folsom_evolution_2018} to derive the stellar parameters of S~CrA~N from the fitting of photospheric lines in our spectra. The \texttt{ZEEMAN} code solves the radiative transfer assuming Local Thermodynamic Equilibrium (LTE) in a 1D stellar atmosphere. 
We used the MARCS stellar atmosphere model and the theoretical properties of the photospheric lines are extracted from the VALD database as mentioned in Sect. \ref{Sec:Data}. 
Then, the procedure for \texttt{ZEEMAN} is to adjust a synthetic spectrum to an observed one by minimizing a $\chi^2$ function taking into account its seven free parameters, which are the effective temperature $T_\text{eff}$, the equatorial rotation velocity of the star projected on the line of sight $v\sin{i}$, the radial velocity $v_r$, the local veiling $r$ taken to be an excess continuum flux as a fraction of continuum, the surface gravity $\log g$, the micro-turbulent velocity $v_\text{mic}$, and the macro-turbulent velocity $v_\text{mac}$. 
A 7D $\chi^2$ map being likely to display several local minima, we adjusted the parameters by pairs: $T_\text{eff}$ along with $r$, then $v\sin{i}$ along with $v_r$, then $v_\text{mic}$ along with $v_\text{mac}$, $\log g$ being adjusted on its own. At first, all constant parameters were set to an arbitrary value until a minimum $\chi^2$ was reached for a considered pair of parameters. Then, the two previously fitted quantities were set constant to their new value, before fitting two new parameters until reaching a new minimum $\chi^2$, which gives two different constant values to the new couple of parameters, and so on until we fitted all parameters. Then, the procedure is repeated until a whole cycle of fitting produces no variation in any of the parameters. We tested the sensitivity to initial conditions by setting different starting values for each parameter, but the procedure always converged to the same set of values, except for $\log g$ and $v_\text{mic}$. Their values did not change, regardless of the number of cycles performed, which shows little sensitivity of our fit to these parameters. For the procedure to run, we fixed $\log g$~=~4.0, which is common for CTTS, and derived the mass and radius as described in Sect.~\ref{Sec:discuss}. Concerning $v_\text{mic}$, we retained a value of 1~km/s minimizing the $\chi^2$ function for the procedure to run, but did not derive any uncertainties based on the $\chi^2$ function.
This fitting is done over 15 spectral windows (the central wavelength of which are mentioned in Fig.~\ref{Fig:veiling}) that display clear photospheric lines (i.e., deeper than 10\% of the continuum level).

All our results are gathered on the first four rows of Table~\ref{table:StParams}. Each value corresponds to the average of the values over the 15 spectral windows, and the uncertainty corresponds to their standard deviation. We obtained an effective temperature of ($T_\text{eff}=4300\pm~100$~K). We retrieved a rotational velocity of $v\sin{i}$~=~10$~\pm$~2~km/s, where $i~=~0^\circ$ corresponds to a face-on object. We obtained a radial velocity of $v_r$~=~0~$\pm$~1~km/s, and a macroscopic turbulence velocity $v_\text{mac}$~=~11~$\pm$~3~km/s.
When plotting the veiling values against the spectral windows in which it is measured, no clear trend is observed unlike in most CTTS where veiling decreases with wavelength in the optical domain \citep[see e.g. ][]{fischer_characterizing_2011}. When plotting the veiling as a function of time for different spectral windows (Fig.~\ref{Fig:veiling}), no clear periodicity can be identified, suggesting that an additional source must be at play, on top of the continuum emission from an accretion spot (e.g. accretion-powered emission lines). The mean veiling $r$ over the 15 spectral windows and over time is as high as 7 on average, and ranges between 4 and 11. 

\subsection{Variability in LSD profiles}

The LSD $I$ profiles presented in Fig.~\ref{Fig:LSD_profiles} show strong variability in intensity, which is unsurprising due to changes in veiling intensity. Indeed, we found a negative correlation between the mean veiling and the equivalent width (EW) of the LSD $I$ profile at each date. Thus, both LSD $I$ and $V$ profiles were scaled to the greatest EW (occurring at the Julian date 2,458,303.93) so they all had the same EW. By doing so, one retains only their intrinsic shape variability presumably due to spots.

The shape of the LSD $I$ profiles changes entirely from one date to the following (Fig.~\ref{Fig:LSD_profiles}). From date 291.98 to 292.95, for instance, the minimum intensity switches from the red to the blue side of the profile, and the small excess observed in the blue wing at first becomes reddened at 292.95. Then at date 303.93, the profile looks just like the one at 291.98 again, with only minor changes in intensity. To try and constrain the origin of this variability, we checked whether the LSD $I$ profiles were deformed differently, depending on the intrinsic depth of the lines included in the LSD mask (see Appendix~\ref{app:spectra} for the detailed procedure). It results that no matter the depth limit of the lines taken into account for the LSD computation, the shapes of the profiles remain the same, with just a loss in signal-to-noise ratio (SNR) when removing more lines. We could not find a way to compute LSD $I$ profiles with attenuated perturbations, and therefore adopted the LSD mask that gives the best SNR in the line profiles.

The source of these distortions must break the spherical symmetry of the surface of the star, and considering the high variability of our LSD $I$ profiles, we cannot interpret them as deviations from a rest profile. 
Hence, we compared the LSD $I$ profiles to a synthetic Voigt profile. We adjusted this profile by fitting it to the median LSD $I$ observed within a velocity range of [-25 km/s; +25 km/s].
This reference profile is shown as a dotted line over the LSD $I$ profiles in Fig.~\ref{Fig:LSD_profiles}.

When looking at the $V$ profiles, they also exhibit a variety of shapes, from flat (date 303.93) to typically anti-symmetric (e.g. at date 292.95). We also note that similar $I$ profiles can display very different $V$ profiles. That is the case for profiles at dates 293.99 and 303.93, where the $I$ profiles are very similar, but the $V$ profiles are respectively strong and flat. Conversely, at dates 291.98, 292.95, and 293.99, the $V$ profiles remain constant in intensity over those 3 observations, with just a smooth drift of the centroid from negative to positive velocities. In contrast, the $I$ profiles are drastically different. There is no evident correlation between the $I$ and $V$ variations, suggesting that brightness inhomogeneities at the surface of S~CrA~N might have additional sources besides brightness spots induced by the magnetic activity.

\begin{figure}[t]
\centering
\includegraphics[ trim = 10 0 107 36, clip, width=0.442\linewidth ]{ 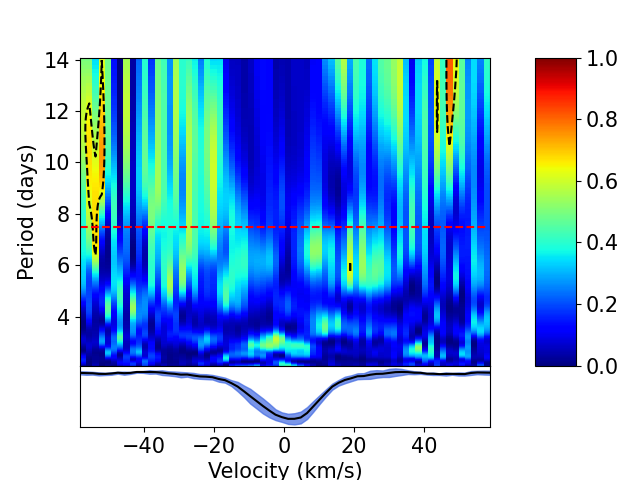 }
\includegraphics[ trim = 55 0 10 36, clip, width=0.508\linewidth ]{ 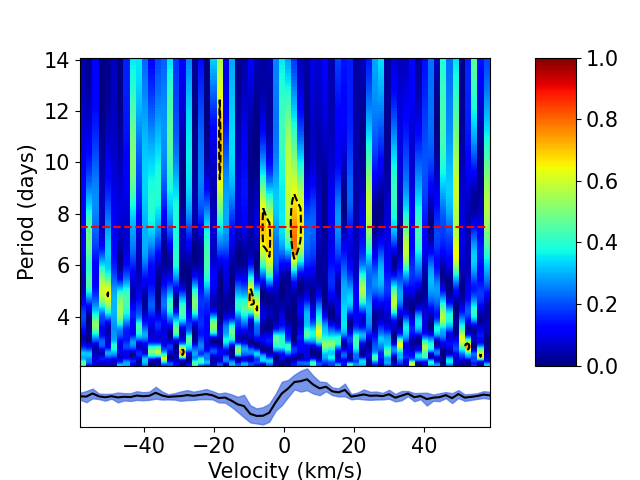 }
\caption{2D periodograms for the LSD $I$ (left) and $V$ (right) profiles. The dashed lines correspond to a period of 7.6 days (red) and FAP~=~3\% contours (black). Below the periodograms are shown the average profiles (black line) surrounded by their 1$\sigma$ deviations (light blue).}
\label{Fig:LSD2DP}
\end{figure}

To recover the stellar rotation period, we computed 2D periodograms for the LSD $I$ and $V$ profiles (Fig.~\ref{Fig:LSD2DP}). We ran a Lomb-Scargle periodogram routine over each velocity channel (i.e., wavelength) to build the periodograms, where a time series of 12 unevenly sampled observations were considered. We restrained the search for periods larger than 2 days (Shannon theorem applied to a 1 day sampling, which is the smallest sampling of our data) and smaller than 14 days. Any periodicity above -or near- 14 days cannot be considered reliable since that was the total span of our observations. False Alarm Probability (FAP) contours of $3\%$ are drawn in Fig.~\ref{Fig:LSD2DP}. All the FAPs of this study were computed assuming white noise in the continuum, i.e. independent measurements, following the method described in \cite{zechmeister_generalised_2009}. The 2D periodograms reveal no apparent periodicity in the LSD $I$ profiles. While these profiles might be influenced by stochastic activity in the post-shock region \citep[see, e.g.][]{petrov_accretion-powered_2011, dodin_interpretation_2012, rei_line-dependent_2018}, which could explain the deformations observed and hide periodic features from hot/cool surface spots, the LSD $V$ profiles are more robust to such a contamination since the intensity of the magnetic field decreases rapidly with distance to the stellar surface ($\propto r^{-3}$ at least, for a pure dipole). 
While the LSD $I$ profiles display no clear periodicity, we can compute, for each line of the LSD $V$ periodogram, the 2D periodogram's powers weighted by a 1-FAP factor over a range of velocity narrowed to the region of variability of the profiles (i.e., [-20~km/s; +20~km/s]). We are left with a 1D weighted periodogram whose peak's maximum is the stellar rotation period and whose standard deviation is the uncertainty. That yields a period of $7.6\pm1.3$~days located at the extrema of the $V$ profile, where the amplitude variation is the most important, strengthening the reliability of this signal detection.

\subsection{Variability in emission lines}

CTTS are known to show strong and broad Balmer emission lines, as well as a narrow emission in He~I lines. They are all often associated with red-shifted absorption features, indicating infall of material \citep{edwards_spectroscopic_1994,beristain_helium_2001}. It is now well accepted that a good part of these lines are formed through magnetospheric accretion \citep{hartmann_magnetospheric_1994, hartmann_accretion_2016}. In our spectra, the Balmer and He I lines show various features, both in emission and absorption, as well as strong variability. We therefore report below our analysis on this variability to understand the origin of formation of these lines, and how they can constrain the magnetospheric accretion processes in a strongly accreting T Tauri star as S~CrA~N.

\begin{figure}[t]
\centering
\includegraphics[trim=0 0 15 35,clip,width=\linewidth]{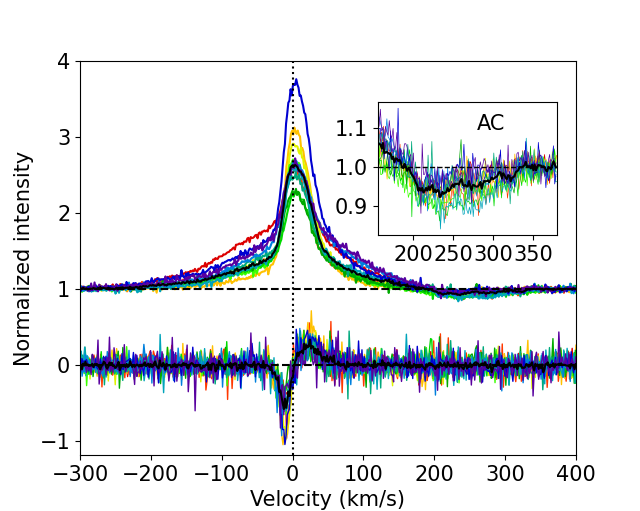}
\caption{He~I~$\lambda$5876 line for all observations. Stokes $I$ (top) is normalised in intensity with respect to the continuum. Stokes $V$ (bottom) is magnified by a factor 5 to make it more visible. The color code is the same as Figure~\ref{Fig:LSD_profiles}. Solid black lines are the average spectra. Dashed black lines represent the continuum level for each component. The dotted black line is the radial velocity of the star. The inset plot is a zoom on the absorption component.}
\label{Fig:HeItot}
\end{figure}

\subsubsection*{\bf Helium I $\boldsymbol{\lambda 5876}$.}
This line can be composite with up to 3 distinct components, which are particularly discernible in the case of S~CrA~N (Fig.~\ref{Fig:HeItot}-top):
\begin{itemize}
    \item A narrow component (NC) which is asymmetric and goes from -30 km/s to 50 km/s and peaks around 0 km/s. Due to its very high excitation potential, this emission line is believed to trace accretion footprints at the surface of the star. 
    \item A broad component (BC), as wide as $\pm$~250 km/s and well reproduced by a Gaussian fit. Its origin is still poorly constrained, but might itself be composite.
    \item An absorption component (AC) in the red wing of the broad component (between $\sim 200$ and $\sim 350$ km/s).
\end{itemize}

Strong Zeeman signatures are present in Stokes V spectra (Fig.~\ref{Fig:HeItot}-bottom). The S-shaped signature spreads from -30~km/s to +50~km/s on average and is as strong as 20\% of the continuum intensity. Due to its broadening and asymmetric shape, this $V$ signal is attributed to the NC and thus traces the magnetic field at the footprint of the accretion columns. In order to study all the components separately, a fit with 3 independent Gaussian profiles was applied to each observed I profile. Each component is extracted by removing the models of the two other components from the profiles: the BC and AC Gaussian models were subtracted to the whole line as they did match the shape and intensity of the observed BC and AC at all phases. On the contrary, due to the intrinsic asymmetry of the NC, removing its Gaussian model left significant NC residuals in the extracted BC and AC profiles. Instead, a smoothed NC was subtracted from the original total profile, along with the AC/BC Gaussian model. This smoothed NC was obtained using a three-point moving average of the extracted NC.

\begin{figure}[ht]
    \centering
    \includegraphics[trim=10 30 25 60,clip,scale=0.45,width=0.85\linewidth]{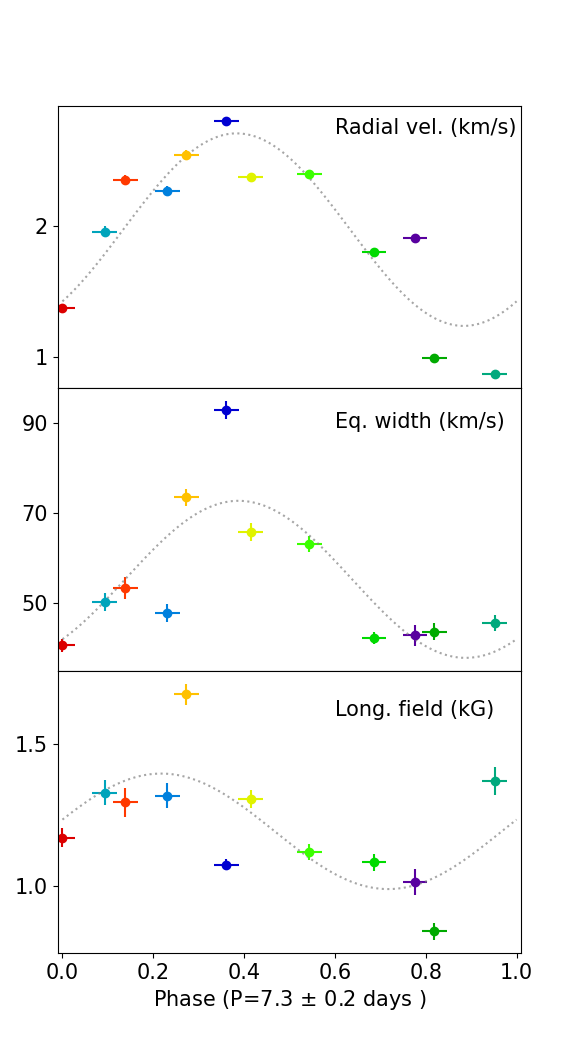}
    \caption{Radial velocities (top), equivalent width (middle), and longitudinal magnetic field (bottom) of the He~I NC as a function of the stellar phase, when considering a stellar rotation period of $7.3\pm0.2$ days and considering the first date of observation as $\phi=0$. When not visible, uncertainties are smaller than the symbol. Dotted black lines illustrate the best fits obtained, with a simple sine (middle and bottom) and the model of \cite{pouilly_beyond_2021} (top). The color code is the same as in Fig.~\ref{Fig:spectra_zoom}. 
    }
    \label{fig:HeI_params}
\end{figure}

We performed radial velocity, equivalent width and longitudinal field measurements in the NC. The longitudinal field is obtained thanks to the first moment method \citep{donati_spectropolarimetric_1997, wade_high-precision_2000} The equivalent width is computed between -30 km/s and +50 km/s. Finally, we measured the radial velocity of the NC's centroid by computing the first moment of the profile at each date. All three quantities are presented in Table~\ref{table:Mag} and Fig.~\ref{fig:HeI_params}. The radial velocity modulation was used to estimate the stellar period thanks to a point-like accretion spot model described in further detail in \cite{pouilly_beyond_2021}. We obtained a period $P=7.3\pm0.2$ days.

\begin{figure}[t]
\centering
\includegraphics[trim=12 0 107 36, clip, width=0.452\linewidth]{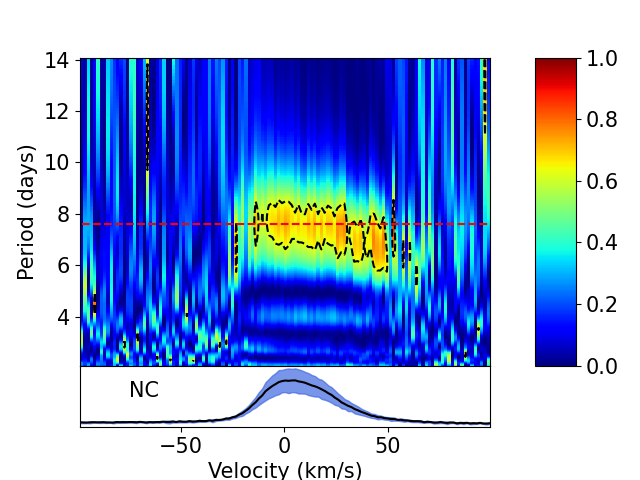}
\includegraphics[trim=52 0 15 36, clip, width=0.52\linewidth]{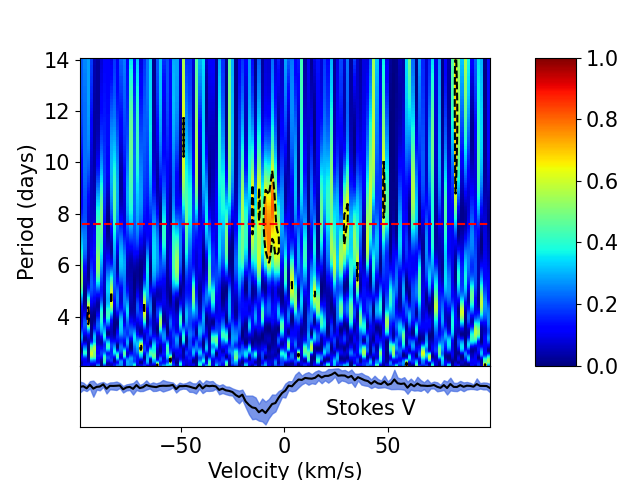}
\includegraphics[trim = 10 0 107 29, clip, width=0.459\linewidth]{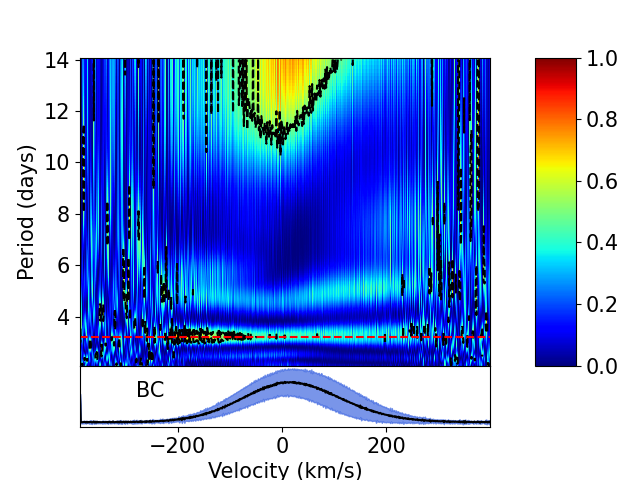}
\includegraphics[trim = 52 0 10 29, clip, width=0.532\linewidth]{ 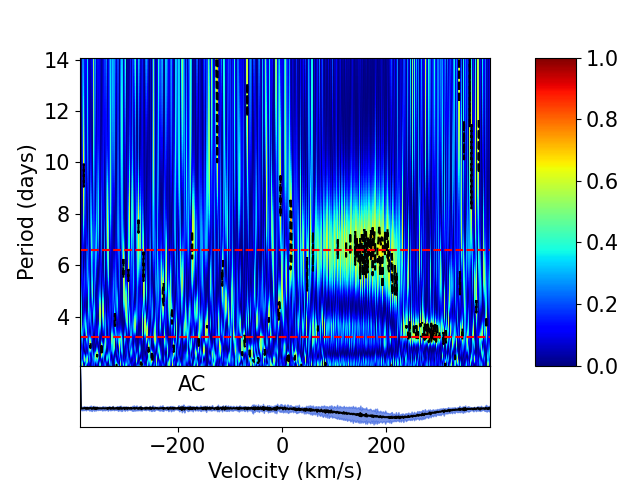 }
\caption{2D Periodograms of the Stokes $I$ NC (upper-left), BC (lower-left), AC (lower-right), and Stokes $V$ (upper-right) in He~I~$\lambda$5876. The red dashed lines mark a period of 7.4 days (NC and Stokes $V$), 3.2 days (BC and AC), and 6.6 days (AC). The black dashed contours denote a constant FAP of 3\% (except 15\% for the BC).
Below the periodograms are shown the average profiles (black line) surrounded by their 1-$\sigma$ deviations (light blue).}
\label{Fig:HeINC2DP}
\end{figure}

\begin{figure}
\centering
\includegraphics[trim=85 12 65 35, clip, width=0.9\linewidth]{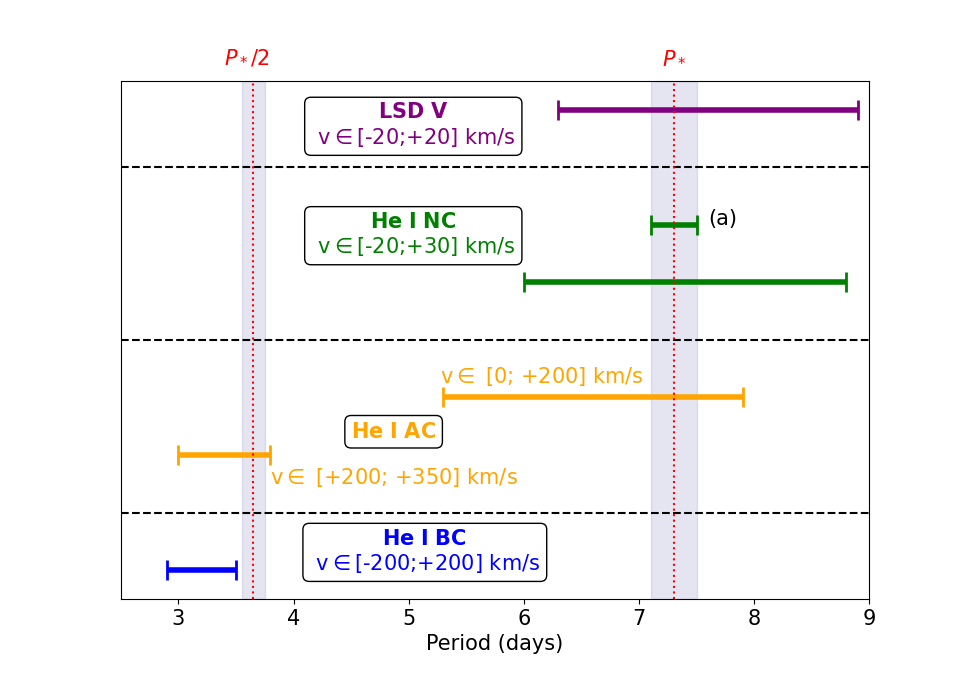}
\caption{Comparison of all the periods derived in this study, gathered according to what profile gave them. Red dashed vertical lines illustrate the stellar rotation period ($P_*$) and half its value, surrounded by their uncertainty (light blue shade). All values are estimated from 2D periodograms except for (a), which is derived from radial velocities, and is chosen as $P_*$. }
\label{Fig:periods}
\end{figure}

The 2D periodograms were computed for $V$ profiles and all three Stokes $I$ components (Fig.~\ref{Fig:HeINC2DP}). Each emitting component displays its own periodicity. The center of the NC shows a 7.4~$\pm$~1.4 days period with strong significance (FAP < $3\%$). However, this period drifts down to 6.6~$\pm$~1.3 days in the red wing (above +30 km/s). The BC shows a 3.2~$\pm$~0.3 days period with a much lower significance (FAP $\sim 15\%$), whereas there is no signal at 7.4 days. The high power signal at long periods cannot be considered significant since it could not be observed for a full cycle, and the white noise assumption in the FAP computations tends to overestimate its significance level. The AC displays a 6.6~$\pm$~1.3 days period from 0 to +200 km/s, and a 3.4~$\pm$~0.4 days period beyond +200 km/s. Both periods are detected with a high significance level (FAP < $3\%$). All the derived periods are gathered and compared in Fig.~\ref{Fig:periods}, and will be discussed in Sec.~\ref{Sec:discuss}.

\begin{table}[ht]
\caption{\label{t7}Radial velocities, equivalent width and longitudinal field derived for each phase in the He~I NC.}
\centering
\begin{tabular}{l c c c}
\hline\hline
Phase & Radial vel. & Equ. width & Long. field  \\
 ($\pm$~0.03) & (km.s$^{-1}$) & (km.s$^{-1}$) & (kG)\\
\hline
0.00 & 1.373~$\pm$~0.003 & 41~$\pm$~1 & 1.17~$\pm$~0.03 \\
0.14 & 2.35~$\pm$~0.04 & 54~$\pm$~2 & 1.30~$\pm$~0.05 \\
0.27 & 2.54~$\pm$~0.03 & 74~$\pm$~2 & 1.68~$\pm$~0.04 \\
0.41 & 2.37~$\pm$~0.03 & 66~$\pm$~2 & 1.31~$\pm$~0.03 \\
0.54 & 2.40~$\pm$~0.03 & 63~$\pm$~2 & 1.12~$\pm$~0.03 \\
0.69 & 1.80~$\pm$~0.02 & 42~$\pm$~1 & 1.08~$\pm$~0.03 \\
0.82 & 0.99~$\pm$~0.01 & 44~$\pm$~2 & 0.84~$\pm$~0.03 \\
0.95 & 0.88~$\pm$~0.01 & 46~$\pm$~2 & 1.37~$\pm$~0.05 \\
1.09 & 1.96~$\pm$~0.04 & 50~$\pm$~2 & 1.33~$\pm$~0.04 \\
1.23 & 2.27~$\pm$~0.04 & 48~$\pm$~2 & 1.32~$\pm$~0.04 \\
1.36 & 2.80~$\pm$~0.02 & 93~$\pm$~2 & 1.08~$\pm$~0.02 \\
1.78 & 1.91~$\pm$~0.03 & 43~$\pm$~2 & 1.01~$\pm$~0.05 \\
\hline
\end{tabular}
\label{table:Mag}
\end{table}

\subsubsection*{\bf Balmer series. }
\begin{figure}[ht]
\centering
\includegraphics[trim=21 8 37 41, clip, width=\linewidth]{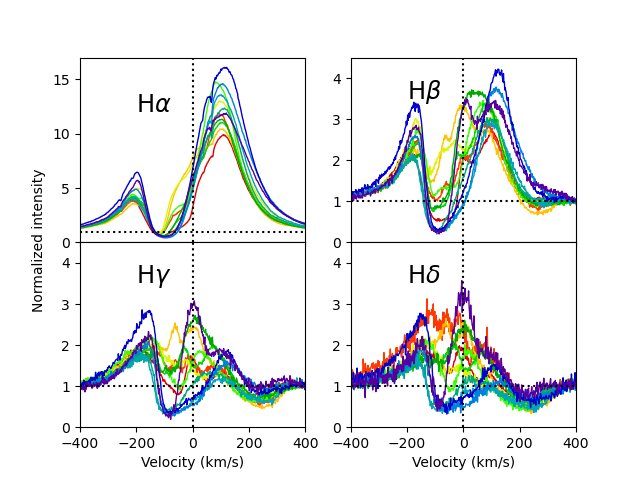}
\caption{Balmer series in S CrA N. The color code is the same as Fig.~\ref{Fig:LSD_profiles}. The vertical dotted line shows the stellar radial velocity, while the horizontal dotted lines show the continuum level for each line. The y-scale of H$\alpha$ has been changed, for clarity.}
\label{Fig:Balmer}
\end{figure}

The detectable Balmer lines in the ESPaDOnS spectral range are displayed in Fig.~\ref{Fig:Balmer}. The corresponding periodograms exhibit signals at periods similar to those found in the LSD and Helium profiles (see Appendix~\ref{app:spectra}). Still, the regions of the lines concerned with these periods are not the same for all the lines, making it hard to interpret which physical process causes their variability. General behaviours can be retrieved nonetheless. The emission lines are broad and asymmetric, ranging from -400 km/s to +400 km/s (even more for H$\alpha$). Albeit variable in intensity and shape, a persistent blue-shifted absorption is present at all phases (peaking at about -100~km/s). On top of this absorption, a narrow emission can be observed in H$\delta$ and H$\gamma$, along with a red-shifted absorption spreading from +50 to +300 km/s. A similar red-shifted absorption is barely detected in H$\beta$ and cannot be seen in  H$\alpha$, although its underlying presence could explain the asymmetry of the far wings in the latter. The exact span of this absorption is not the same for all the lines though. The complexity of the Balmer lines can be interpreted as a multi-component origin of the hydrogen emission, coming from different phenomena and/or different locations in the inner disk and magnetospheric regions, which will be discussed in Section~\ref{Sec:discuss}.

\begin{figure*}[t]
\centering
\includegraphics[trim=20 18 17 16, clip, width=0.32\linewidth]{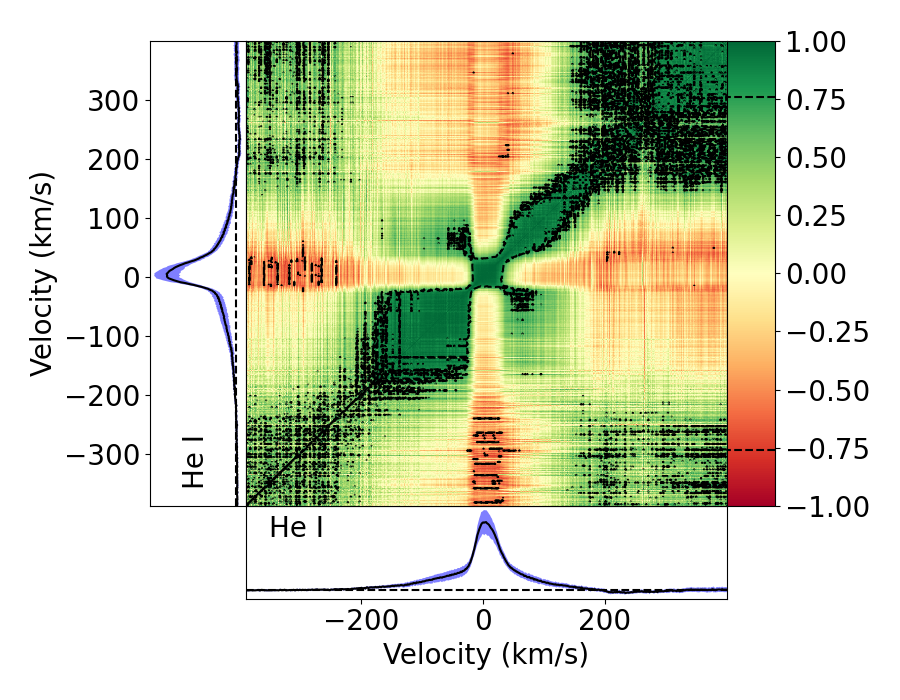}
\includegraphics[trim=20 18 17 16, clip, width=0.32\linewidth]{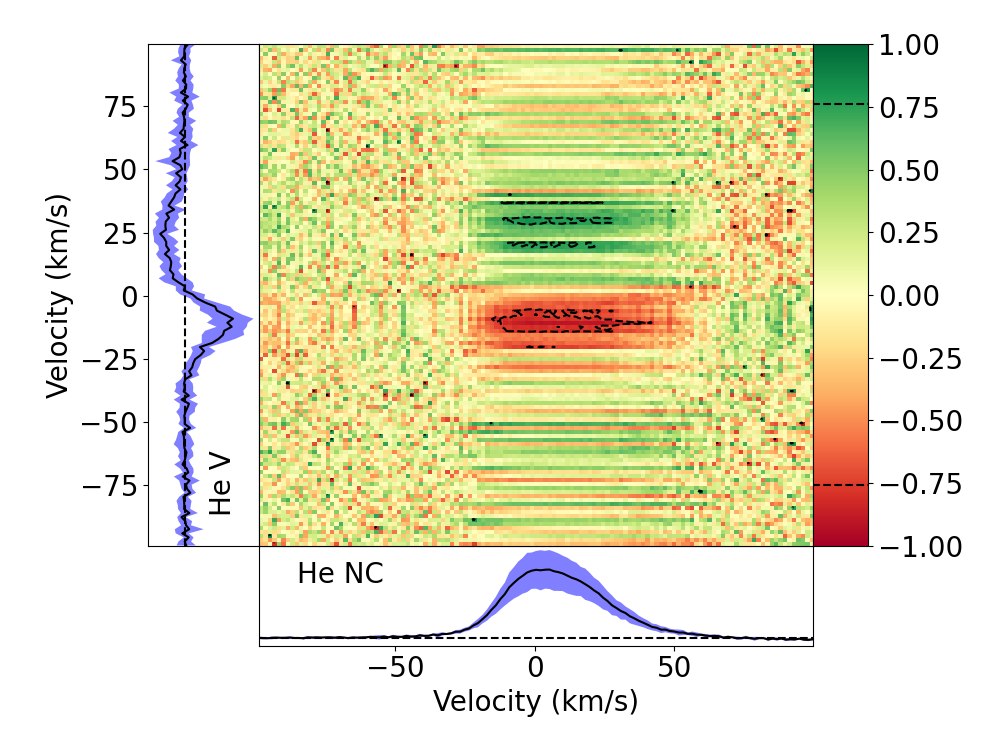}
\includegraphics[trim=20 18 17 16, clip, width=0.32\linewidth]{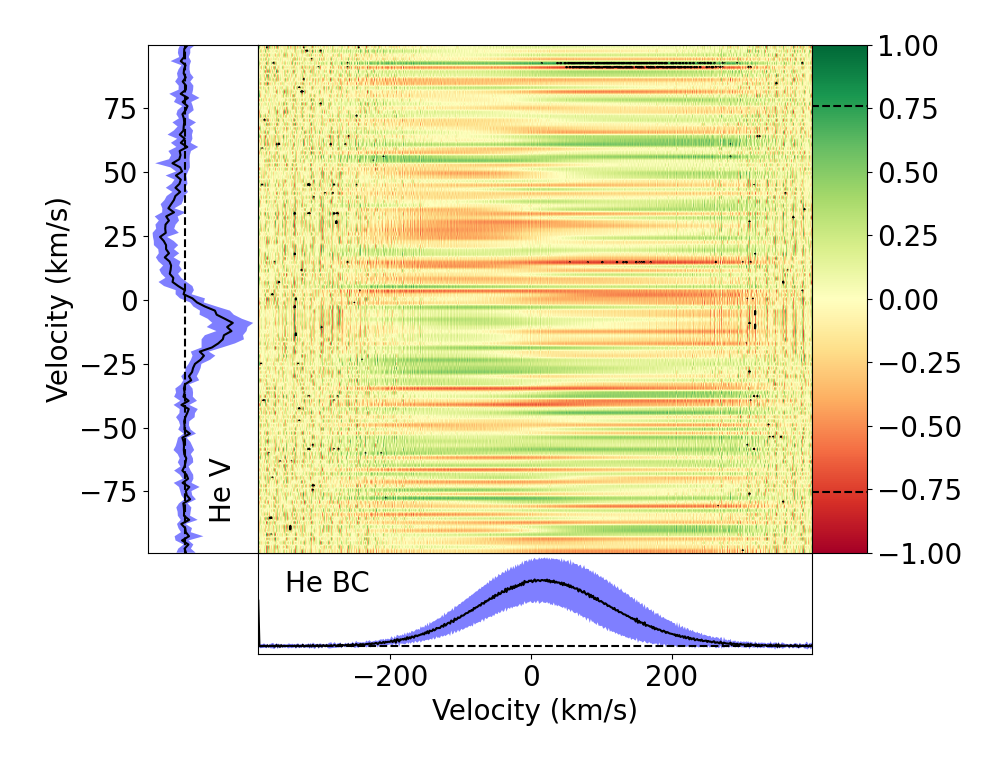}
\includegraphics[trim=20 18 17 16, clip, width=0.32\linewidth]{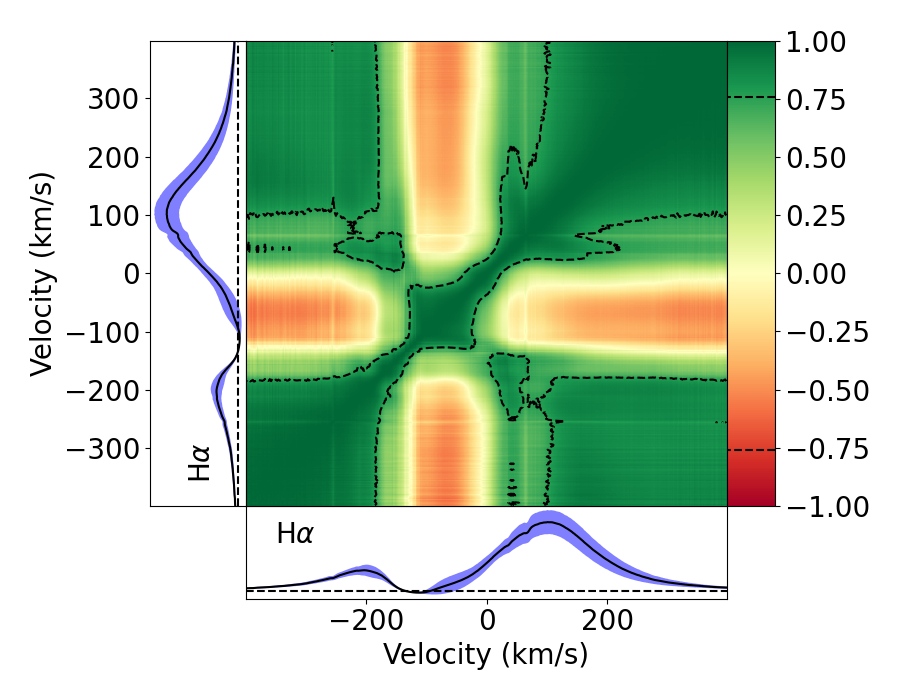}
\includegraphics[trim=20 18 17 16, clip, width=0.32\linewidth]{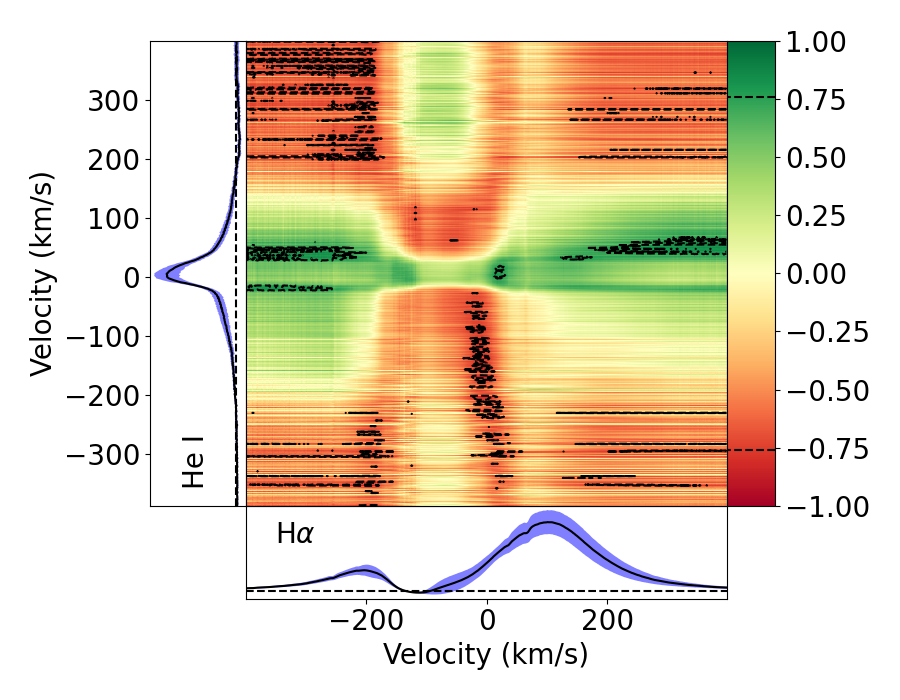}
\includegraphics[trim=20 18 17 16, clip, width=0.32\linewidth]{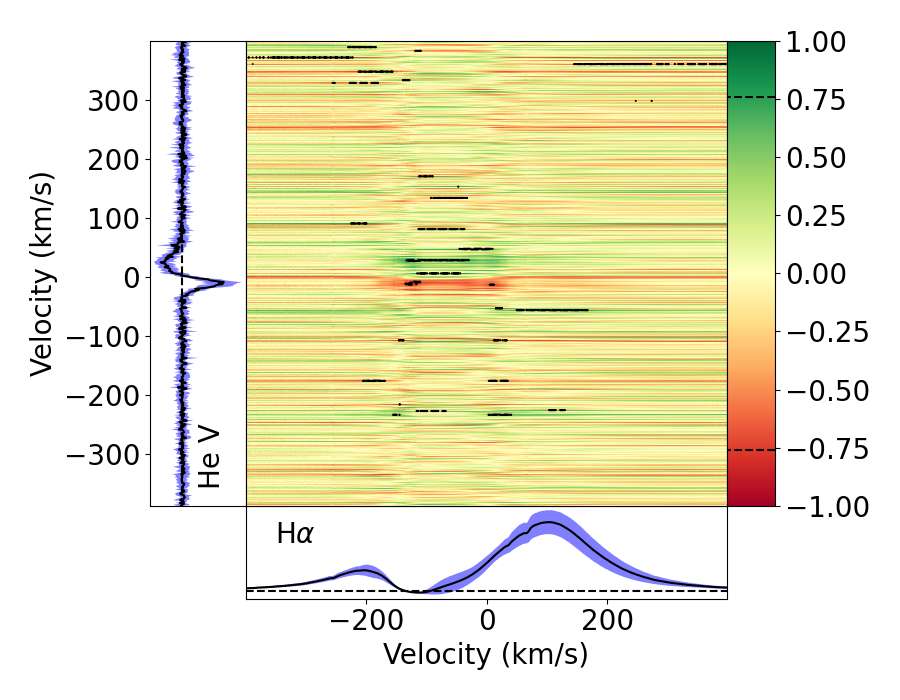}
\caption{Correlation matrices in He~I (top row): auto-correlation of the full Helium line at 587.6~nm (left), correlation between the He~I NC and the $V$ profile (middle), and correlation between the He~I BC and the $V$ profile (right). Correlations matrices in H$\alpha$ (bottom row): auto-correlation of the full H$\alpha$ line (left), correlation between the H$\alpha$ line and the full He~I line (middle), and correlation between the H$\alpha$ line and the He~I $V$ profile (right). The dashed black contours show the $2\sigma$ significance level. The average profiles are represented on the bottom and left side of each panel (solid black line) and are surrounded by their standard deviation (light blue shade), with the continuum level shown with dashed black lines.}
\label{Fig:Corr}
\end{figure*}

\subsection{Correlations in spectral lines}

We computed correlation matrices based on the Pearson coefficients to highlight different behaviours within a line profile and/or to link different lines between them \citep[see e.g. :][]{kurosawa_time-series_2005, pouilly_magnetospheric_2020}. More explicitly, for two given lines, one can compute the correlation between each pixel (i.e., velocity channel or wavelength) of the first line and each pixel of the second line, thanks to the time series of these lines. 
Given a pixel $i$ in the first line and a pixel $j$ in the second line, the correlation coefficient $R_{ij}$ between these pixels is given by:
\begin{equation}
    R_{ij}=\frac{C_{ij}}{\sqrt{C_{ii}C_{jj}}}
\end{equation}
with $C_{ij}$ the covariance between $i$ and $j$ : 
\begin{equation}
    C_{ij}=\frac{1}{N-1}\sum_{k=1}^N \left(F_{i,k}-\overline{F_i}\right)\left(F_{j,k}-\overline{F_j}\right)
\end{equation}
where $N$ is the number of observations (12 here), $F_{i,k}$ is the flux in pixel $i$ for observation $k$, and $\overline{F_i}$ is the mean flux in pixel $i$ over the $N$ observations.\\
The coefficients of $R_{ij}$ hence range from -1 (perfect anti-correlation) to +1 (perfect correlation), with a null value meaning no correlation. To define some value above which we would consider a correlation level as significant, we computed 1 billion samples where two random variables were taken 12 times (one for each observation) in a white noise equivalent to our Stokes $I$ RMS~=~0.05. This gives a Gaussian distribution of $R_{ij}$ centered on 0 and with $\sigma=0.38$. We chose to take a 2$\sigma$ level of significance, corresponding to $\lvert R_{ij}\rvert \geq 0.76$ for significant correlation, while $0.38 \leq \lvert R_{ij}\rvert \leq 0.76$ will correspond to a moderate correlation, and $\lvert R_{ij}\rvert \leq 0.38$ to low correlation.

\subsubsection*{\bf Correlations in He~I. }
The decomposition of the Helium line into BC and NC components is well justified by the auto-correlation matrix of the whole line (Fig.~\ref{Fig:Corr}-top-left). These components are neither correlated nor anti-correlated. Hence, their origin is linked to different processes. When checking for the correlations between each component and the Stokes $V$ signature, the NC is well correlated with the positive peak of the $V$ profile, while anti-correlated with the negative peak. The BC displays no clear correlation with any part of the $V$ profile, confirming that the $V$ signal is produced by the magnetic field in the region of formation of the NC, and is unrelated with the BC formation.

\subsubsection*{\bf Correlations in $\boldsymbol{H\alpha}$. } 
Fig.~\ref{Fig:Corr}-bottom-left shows the auto-correlation matrix of the H$\alpha$ line. Two main components can be identified: a main broad symmetric emission (from -400 to 400 km/s), truncated by a strong absorption from -200 to 0 km/s. It should be noted that this blue-shifted absorption is actually twofold: a saturated part spreads between -200 and -100 km/s, and a variable one between -100 and 0 km/s. This variable component is moderately anti-correlated ($R_{ij}<-0.38$) with the broad emission.

\subsubsection*{\bf Cross-correlations between He~I and $\boldsymbol{H\alpha}$.}
All the correlations found between the H$\alpha$ and the He~I lines are moderate compared to the ones found for the previously mentioned auto-correlations matrices. Still, hints for possible correlation are visible and presented here. Fig.~\ref{Fig:Corr}-bottom-middle shows that the He~I NC seems correlated with the saturated part of the blue-shifted absorption in the H$\alpha$ line, even though the correlation coefficient does not reach the 2$\sigma$ significance level. This partial correlation is confirmed when inspecting Fig.~\ref{Fig:Corr}-bottom-right. Indeed, the $V$ profile exhibits the same correlation pattern with both the variable part in the blue-shifted absorption of H$\alpha$ and the He~I NC (see Fig.~\ref{Fig:Corr}-top-middle). A moderate anti-correlation ($R_{ij}\sim -0.38$) is found between the variable blue-shifted absorption of H$\alpha$ and the He~I BC. However, considering the previous results, a negative correlation between the He~I~V profile and the He~I BC should be found (Fig.~\ref{Fig:Corr}-top-right), which is not the case. Either there is no correlation between these components, or the underlying correlation is  being quenched due to a potential cross-talk.
Such a moderate correlation can also be found between the whole blue-shifted absorption of H$\alpha$ and the He~I AC, with $R_{ij}\sim 0.38$ (green light area at the top of Fig.~\ref{Fig:Corr}-bottom-middle). Only this time, we do not have other diagnoses to confirm this correlation.

\subsection{Magnetic field reconstruction}

\begin{figure}[ht]
\centering
\includegraphics[trim= 58 77 85 140, clip, width=0.95\linewidth]{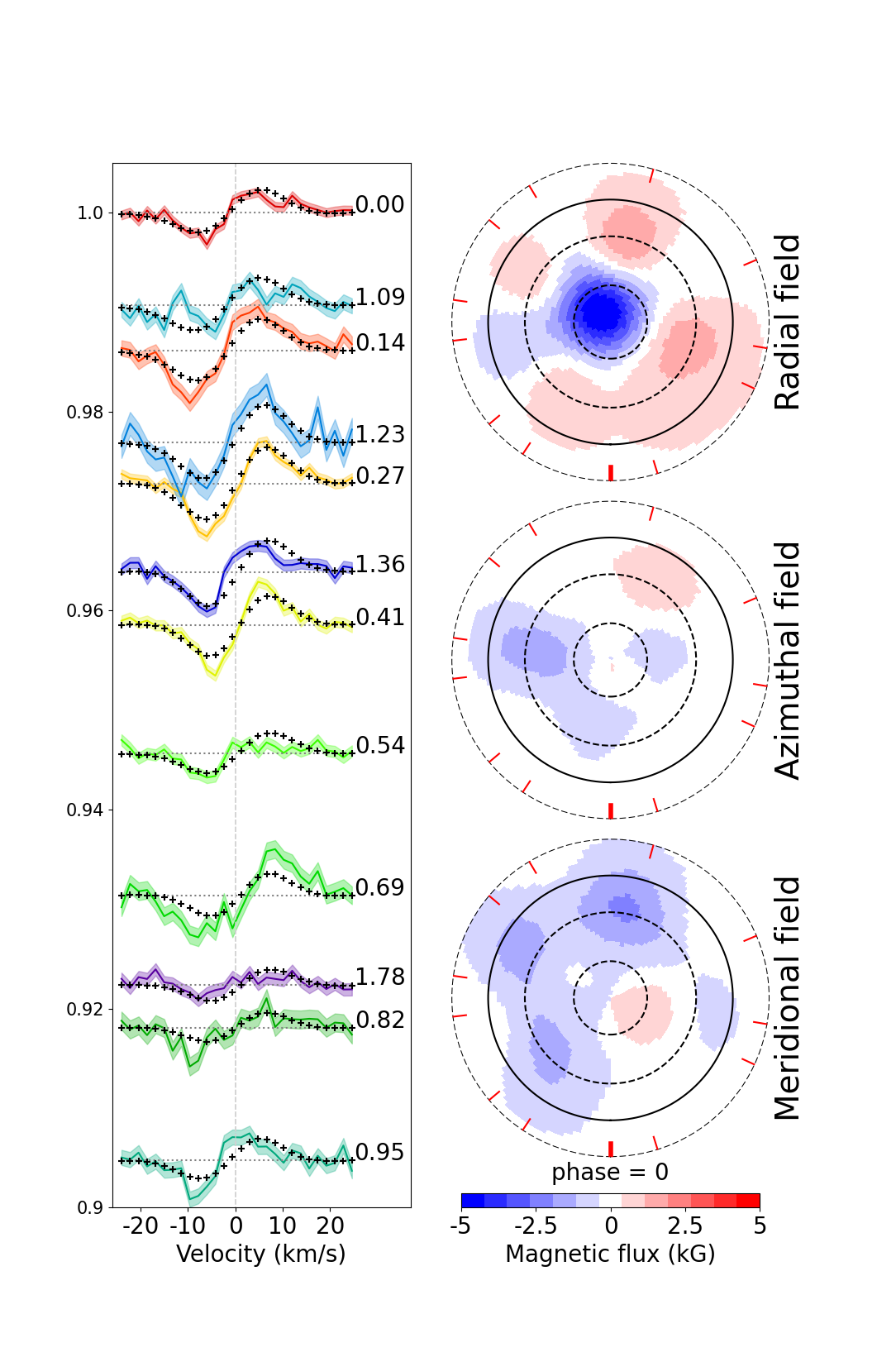}
\caption{Polar projection of the magnetic maps (right) reconstructed from the LSD $V$ observations (left, coloured lines). Black crosses overplotted to the LSD profiles show the reconstructed $V$, on the right of which the cycle number and phase are mentioned. The red ticks surrounding each map show the observed phases, going increasingly clock-wise, starting from the South direction. Dashed black circles show colatitudes of $27^\circ$ (i), $63^\circ$ (90-i) and $117^\circ$ (90+i). The solid black line shows the equator.}
\label{Fig:MagMaps}
\end{figure}

The Zeeman Doppler Imaging (ZDI) technique uses the rotational modulation of Stokes $V$ profiles to infer the large-scale magnetic configuration. We use the \texttt{ZDIpy} code which assumes a decomposition of the field in spherical harmonics and adopts a weak-field approximation to reconstruct it's topology, as described in full details in \cite{folsom_evolution_2018}.
\texttt{ZDIpy} applies a regularized fitting algorithm that simultaneously maximizes the entropy of the reconstructed topology while keeping the corresponding $\chi^2$ below some target value \citep{skilling_maximum_1984}. The obtained map is therefore interpreted as the minimal information map fitting the observed LSD $V$ data with a goodness determined by the target $\chi^2$.

The ZDI procedure described above usually also fits Stokes $I$ profiles to reconstruct the brightness distribution of the star. However, for S~CrA~N, we never reached a satisfying fit for the LSD $I$. That is why a uniform brightness was assumed. This uniform brightness was obtained with a Voigt model fitting the observed median shape of LSD $I$ (see Fig.~\ref{Fig:LSD_profiles}). In order to reconstruct the magnetic maps, one needs to know the stellar rotation period and inclination with respect to the line of sight. From the variability analysis of the He~I line, we adopt a stellar period $P_*=7.3\pm0.2$ days (see Table~\ref{table:StParams}). Combining this stellar period with $v\sin{i}$, and a radius estimate presented in Section~\ref{Sec:discuss}, we derive the star's inclination $i=39^\circ \pm 16^\circ$, which is in agreement at a 1-$\sigma$ level with the inclination of the inner disk as determined by \cite{gravity_collaboration_gravity_2021} (27~$^\circ \pm$~3$^\circ$). Considering the propagation of uncertainties in the computation of our stellar inclination, we adopt hereafter the value from \cite{gravity_collaboration_gravity_2021} as the inclination of the star: $i=27^\circ \pm 3^\circ$. The ZDI procedure ran over 16 iterations to maximize entropy reaching the target reduced $\chi^2$ of 1.5. This value represents the smallest target we could reach without fitting the noise of the data. The spherical harmonics expansion was truncated to consider only the modes where $l\leq 10$ and reproduced the LSD $V$ profiles between -25 km/s and +25 km/s.\\

The reconstructed $V$ profiles are over-plotted to the observed LSD $V$ profiles along with the derived magnetic maps in Fig.~\ref{Fig:MagMaps}. The three maps represent the projection of the magnetic vector on the spherical basis, which translates into a radial field (positive polarity meaning a vector going out of the surface), azimuthal field (positive polarity meaning a vector oriented clock-wise), and meridional field (positive polarity meaning a vector oriented toward the South direction). We represent on each map the rotation phases at which each observation is obtained, with the initial phase $\phi=0$ corresponding to the first date of observation. The large-scale magnetic field best reproducing our Stokes V profiles is as strong as 5.4~kG (which is within the weak-field approximation validity domain), with a mean intensity of 950~G, and a global axisymmetry of 68\%. The poloidal component of this field (82\% of the total energy) is primarily located in the dipolar component, which represents 27\% of the total magnetic energy, even though higher order poloidal modes are significantly present (up to 25\% and 16\% of the total magnetic energy for the quadrupole and the octupole, respectively). This dipole's maximum intensity (816~G) is reached at intermediate latitude ($56^\circ$), at phase 0.29. The main magnetic properties of the system are listed in Table~\ref{table:MagProps}.
\begin{table}[ht]
\caption{Main magnetic field properties from ZDI.}
\centering
\begin{tabular}{lc}
\hline\hline
Property & Value  \\
\hline
Maximum intensity & 5.4 kG  \\
Mean intensity & 950 G  \\
Axisymmetry\tablefootmark{a} & 68 \% \\
\hline
\textbf{Toroidal field}\tablefootmark{a}  & \textbf{18\%} \\
\hspace{4 mm}Axisymmetry\tablefootmark{b} & 41 \% \\
\hline
\textbf{Poloidal field}\tablefootmark{a}  & \textbf{82\%} \\
\hspace{4 mm}Axisymmetry\tablefootmark{c} & 73 \% \\
\hspace{4 mm}Dipole\tablefootmark{c} & 33 \% \\
\hspace{7 mm} Maximum intensity & 816 G \\
\hspace{7 mm} Pole's location\tablefootmark{d} [lat; phase] & [56$^\circ$; 0.29] \\
\hspace{4 mm}Quadrupole\tablefootmark{c} & 30 \% \\
\hspace{4 mm}Octupole\tablefootmark{c} & 19 \% \\
\hline
\end{tabular}
\tablefoot{
\tablefoottext{a}{expressed in \% of the total magnetic energy}; \tablefoottext{b}{expressed in \% of the toroidal field's magnetic energy}; \tablefoottext{c}{expressed in \% of the poloidal field's magnetic energy}; \tablefoottext{d}{Visible pole, negative polarity.}
}
\label{table:MagProps}
\end{table}

\section{Discussion}
\label{Sec:discuss}

\subsection{S CrA N: a strong accretor ?}

Typical CTTS have effective temperatures between 3000 and 5000\,K, (spectral type G or later), masses most generally between 0.3 and 1\,\msun, with an upper limit of about 2\,\msun, and mass accretion rates of the order of $10^{-7}$-$10^{-10}$\,\msunpyr.
They probe the end of the fully convective Hayashi phase of the pre-main sequence evolution, around ages of 1 to 2\,Myr \citep[e.g.][]{herczeg_optical_2014,villebrun_magnetic_2019, nicholson_surface_2021}.

When fitting our ESPaDOnS spectra, we found an effective temperature for S~CrA~N of 4300~$\pm$~100\,K, in agreement with the previous determinations (Table~\ref{table:StParams}). We used the Gaia DR3 data and applied the method of \cite{galli_corona-australis_2020} to derive the distance to S~CrA~N (see Appendix~\ref{app:dist}). We adopted a value of d~=~$152.4 \pm 0.4$ pc. 
We used the luminosity $L_*=1.67\pm0.8~L_\sun$ from \cite{prato_astrophysics_2003}, corrected with our new distance to place S~CrA~N in the Hertzsprung-Russell diagram. Using the CESAM evolutionary model \citep{morel_cesam_2008, marques_seismic_2013}, we estimated a mass $M_*=0.8\pm0.1$~\msun\ and stellar radius $R_*=2.3\pm 0.6~R_\sun$, which gives $\log g = 3.6\pm0.2$. The star is placed at an age of about 1\,Myr, i.e. younger than typical CTTS. As most of the CTTS, S~CrA~N is then fully convective.

The photospheric lines of CTTS generally appear shallower than those of non-accreting stars, suggesting an excess of continuum emission. This so-called veiling can be observed in different regions of the spectrum of CTTS, with different interpretations: in the near-infrared range, this excess might be attributed to the dust emission of the protoplanetary disk \citep{sousa_new_2023}; in the ultraviolet-visible range, this excess is attributed to an accretion shock that behaves like a $\sim$ 8\,000~K blackbody continuum emission that superimposes to the photospheric continuum. In very active stars, emission lines spectra may also blend the photospheric lines as early discussed in, e.g., \cite{bertout_review_1984} and shown in, e.g., \cite{petrov_accretion-powered_2011, dodin_interpretation_2012, rei_line-dependent_2018}. This veiling of optical spectral lines is typically lower than 2 around 5500~\AA~\citep[]{basri_hamilton_1990,hartigan_optical_1991}. In S~CrA~N, we found no trend between wavelength and the veiling, which happens to be strongly variable around 5500~\AA\ with values ranging from 2 to 11. This finding is consistent with that of \cite{sullivan_s_2019}, who also detected a strong variability and a large amplitude (between 2-6) of veiling in the near-infrared range. Combined with its young age, this evidence of very strong accretion could indicate an evolutionary stage between Class I and Class II for S~CrA~N. This is further confirmed by the place of the S~CrA system in the color-color diagram proposed by \cite{koenig_classification_2014}. With All-WISE measurements $W1=5.1\pm0.1$, $W2=4.0\pm0.1$ and $W3=2.08\pm0.01$, the binary system lands at the frontier between protostars and T Tauri stars\footnote{\footnotesize With All-WISE, S~CrA~N is observed together with S~CrA~S, due to the angular resolution of the observations ($\sim 6"$). Since the two components are coeval \citep{gahm_s_2018}, this does not affect the interpretation regarding their young age.}.

\subsection{The magnetosphere}
\label{Sec:the magnetosphere}

From its fundamental parameters, S~CrA~N can be pictured as a young, fully convective T Tauri star. Should the surface magnetic field of CTTS be controlled by their internal structure only, we might expect a strong ($\sim$1~kG) axisymmetric large-scale field, mostly poloidal, with a strong dipole relative to higher poloidal modes \citep{gregory_can_2012}. This expectation is only partially met by the magnetic maps obtained in Section~\ref{Sec:Results}: the total field is mostly (82\%) a poloidal field that is axisymmetric (73\%), with a strong (816~G) dipole. However, the dipole and the quadrupole represent a similar portion of the poloidal field (33\% and 30\%, respectively), while the octupole's contribution is smaller, but still significant (19\%).
Our uncertainties on the luminosity may be the reason for this difference between the prediction and the reconstructed maps, but it should be noted that this region of the H-R diagram is observationally found to be populated with a variety of magnetic topologies \citep{donati_magnetic_2020, nicholson_surface_2021}. It is also worth reminding that S~CrA~N is a strong accretor, and that the possible impact of accretion on the inner structure of stars is still to be explored.

We estimate the truncation radius (i.e. radius at which the magnetic pressure equals the ram gas pressure) from this magnetic reconstruction, using the formula from \cite{bessolaz_accretion_2008}:
\begin{equation}
    \frac{R_t}{R_*} \simeq 2\,B_*^{4/7}\Dot{M}^{-2/7}M_*^{-1/7}R_*^{5/7},
    \label{Eq:Bessolaz}
\end{equation}
where the stellar field in the equatorial plane $B_*$ (i.e., half the dipole's maximum intensity) has been normalised to 140 G, the accretion rate $\Dot{M}$ to $10^{-8}$ \msunpyr, the stellar mass to 0.8 \msun, and the stellar radius to 2 \rsun, and where a Mach number of $\sim 1$ has been assumed. Since the ESPaDOnS data are not flux-calibrated, we estimated the accretion rate thanks to the width of H$\alpha$ at 10\% peak intensity using the relationship from \cite{natta_accretion_2004}:
\begin{equation}
    \log{\Dot{M}_\text{acc}} = -12.89 (\pm 0.3) + 9.7(\pm 0.7)\times 10^{-3}\; \text{H}\alpha 10\%,
\end{equation}
where $\text{H}\alpha 10\%$ is the considered width in km/s, and $\Dot{M}_\text{acc}$ is in \msunpyr. With widths ranging from 579 km/s to 693 km/s, we obtained accretion rate logarithms between -7.3 and -6.2 with a median value $\log{\Dot{M}_\text{acc}}~=~-6.9 \pm 0.7$, which is consistent with the value of $(1.0 \pm 0.1)\times10^{-7}~$~\msunpyr~from \cite{sullivan_s_2019}, when correcting the distance. Because they use the Br$\gamma$ line to derive this accretion rate, we considered this value as more reliable than our own \citep[see][]{alcala_x-shooter_2014} to compute the truncation radius. Combined with the stellar radius and mass, and the magnetic topology from the present work, we obtained:
\begin{equation}
    R_t = 2.1 \pm 0.4~R_*,
\end{equation}
where the uncertainty is derived from the propagation of errors on the involved parameters through Eq.~\ref{Eq:Bessolaz}. This value is likely a lower boundary for the truncation radius since the ZDI technique is blind to the magnetic field in the region of formation of the emission lines, leading to inconsistency between our reconstructed maps and the direct diagnoses of accretion, as shown below.

The derived truncation radius implies a colatitude of the accretion spot of 44$^\circ$~$\pm$~5$^\circ$ in a purely dipolar accretion model, which is consistent with the derived obliquity of the dipole from ZDI (56$^\circ$). However, both values seem to contradict the very high apparent latitude of the post shock region as traced by the He~I NC radial velocity curve (Fig.~\ref{fig:HeI_params}-top). The single spot model from \cite{pouilly_beyond_2021} applied to these data gives a latitude of 86$^\circ$~$\pm$~1$^\circ$. The asymmetry of the NC attributed to a velocity gradient in the post-shock region does not vary with phase, which is also supportive of a spot located at high latitude.

The magnetic field projected along the line of sight (or longitudinal magnetic field $B_l$) associated with the NC of He~I is positive at all phases (see Fig.\ref{fig:HeI_params}-bottom, Table~\ref{table:Mag}). If one assumes that the NC is arising from the post-shock region, this means at least part of the total magnetic field should be positive near the pole of the star, as shown earlier. However, the magnetic maps reconstructed from ZDI (Fig.~\ref{Fig:MagMaps}) show that regions with a positive magnetic field (i.e. coming towards the observer) at virtually every phase are only present in colatitudes larger than $\sim 60^\circ$, which corresponds to a portion of the star occasionally visible only (outside the second dashed black ring). This discrepancy leads us to consider the tomographic reconstructions presented in this work as not definite.
Additional developments considering this emission line or any other line constraining a different region than the LSD profiles (e.g., Ca~II infrared triplet (IRT), Fe~II 42 multiplet) are required to combine the different constraints in a single tomographic reconstruction (such as in \citealp{donati_magnetic_2020}), which is beyond the scope of this paper.

Finally, from this truncation radius, one can also deduce the expected accretion velocity $v_\text{acc}$ (i.e. the velocity of the gas free-falling from the truncation radius onto the stellar surface) thanks to energy conservation. This gas is producing the AC observed in He~I (Fig.~\ref{Fig:HeItot}), the higher members of the Balmer series (Fig.~\ref{Fig:Balmer}) and FeII (Fig.~\ref{Fig:emission_lines}). Thanks to our estimate of stellar mass and radius, we obtain $v_\text{acc}$~=~267~km.s$^\text{-1}$. This velocity is included in the range of velocities of the AC, which makes the value of $R_t$ derived from ZDI consistent with our values of $M_*$ and $R_*$.
However, the maximum velocity of the AC is larger than the free-fall velocity from infinity ($v_\infty$~=~370~km.s$^\text{-1}$), which points out either a wrong estimate of $M_*$ and $R_*$, or a significant broadening of these absorptions, the source of which remaining unknown.

\subsection{An unstable accretion regime ?}

The observed intensity and variability of various emission lines (He~I, Fe~II, Ca~II, for instance; see Appendix~\ref{app:spectra}), combined with the highly veiled photospheric lines, suggest that intense accretion processes occur in the vicinity of S~CrA~N.

One way to better constrain the accretion process is to compare the corotation radius and the truncation radius. Indeed, when the truncation to corotation radii ratio becomes small enough, Rayleigh-Taylor instability can be triggered at the magnetosphere-disk boundary \citep{kulkarni_accretion_2008}. Then, magnetospheric accretion enters the so-called unstable accretion regime, where classical accretion funnels are observed, but also accretion tongues. This phenomenon and its observable signatures have been extensively modeled (\citealp{kurosawa_spectral_2013}; \citealp{blinova_boundary_2016}) and has recently been reproduced in laboratory experiments that scale to the expected YSOs' accretion tongues \citep{burdonov_laboratory_2022}.
\citet{blinova_boundary_2016} observed unstable accretion for simulations where $R_t/R_\text{co}<0.71$, for obliquities lower than $20^\circ$.\\
With the set of stellar parameters we derived (see Table~\ref{table:StParams}) we computed a corotation radius of $R_\text{co}=6.4\pm1.7$~R$_*$, which yields a truncation to corotation radii ratio $R_t/R_\text{co}=0.33~\pm~0.11$, and places S~CrA~N in the unstable accretion scenario. The simulations from \cite{kurosawa_spectral_2013} showed that a proxy to the accretion regime lies in the profiles of the higher members of the Balmer series ($H\gamma$ and $H\delta$): in the stable case, a red-shifted absorption appears for about half the rotation cycle, before it disappears, coming and going with stellar periodicity. This absorption is due to the visible accretion funnel which is absorbing the star's continuum  while it remains in the line of sight. In the unstable case, this red-shifted absorption is expected to be present at virtually every phase of the rotation cycle, since there is at least one accretion tongue in the line of sight at all time. The profiles of $H\gamma$ and $H\delta$ we observe in S~CrA~N display this absorption at all phases as seen in Fig.~\ref{Fig:Balmer}, in agreement with unstable accretion.

\subsection{Accreting structures}

This unstable accretion regime sets a new context for the interpretation of the features observed in the emission lines presented earlier.

The He~I $\lambda 5876$ NC likely arises from a post-shock region in a single accretion spot located at high latitudes ($\sim86^\circ$). The origin in a post-shock region is suggested by the small radial velocity of the flow producing the NC ($\sim2$~km/s, see Sec.\ref{Sec:the magnetosphere}). The single-spot model is corroborated by the consistency between the periodicity of the Stokes V signatures (both in the LSD profiles and the He~I line) and the periodicity of the He~I NC profiles.
The longitudinal field's curve in the He~I line displays a complex modulation, with a substantial departure from the sine model near the extreme radial velocities, around phases 0.3 and 0.9 (see Fig.~\ref{fig:HeI_params}). That could be explained if the star has a complex magnetic field, significantly differing from a simple dipole as suggested by ZDI from the present work. Finally, the asymmetry of the NC in Stokes $I$ (steep blue wing and mild red wing; see Fig.~\ref{Fig:HeItot}) and the departure from anti-symmetry in Stokes $V$ (blue lobe stronger than red lobe) would come from the velocity gradient in the region of emission. This asymmetry is observed at all phases with no noticeable modulation, suggesting a high latitude for the accretion spot.

We speculate that the red-shifted AC seen in He~I $\lambda 5876$, $H\gamma$, and $H\delta$ is formed in two distinct structures: first, in an accretion funnel located around phase 0.2 (called the "main" column hereafter), which produces the NC at its footprint, when forming an accretion shock; then, in another accretion column located around phase 0.7 (called the "secondary" column hereafter, opposed to the main one), which does not have any NC counterpart, because its density is much smaller than in the main column, where accretion is favoured, due to the magnetic misalignment. An AC is nonetheless produced by the free-falling material of the secondary column which absorbs the light emerging from the photosphere. These opposite structures then produce an artificial periodicity of half the stellar period ($P_*/2=3.65\simeq3.4\pm0.4$~days, see Fig.~\ref{Fig:periods}), the actual stellar period being detected for the lowest velocities of the AC ($P_*=7.3\simeq6.6\pm1.3$~days), because only the main accretion column is dense enough at small velocities (i.e., in the upper part of the column) to produce an absorption.

Finally, the He~I $\lambda 5876$ BC is most likely formed by the in-falling gas in both the main and the secondary funnels, since the same artificial periodicity as the AC is observed, and its centroid is mostly red-shifted. As discussed in \cite{beristain_helium_2001}, this tendency in the central velocity of the line can be obtained with material following a purely dipolar field at polar angles lower than 54.7$^\circ$ which corresponds to accreting gas.

\subsection{Outflows}
\cite{lima_modeling_2010} modeled the H$\alpha$ line in the case of a CTTS undergoing both magnetospheric accretion and a disk wind. When comparing the H$\alpha$ lines we observed (Fig.~\ref{Fig:Balmer}) with their grids of models, the deep blue-shifted absorption of our profiles appears to be well reproduced with their models combining a high accretion rate ($10^{-7}$\,\msunpyr), and a hot ($\sim$ 8000-10000~K) disk wind with a density of a few $10^{-11}$~g/cm$^3$ and an outer radius of a few tens of stellar radii (see their Figs. 4 and 6). It should be noted that their model reproduces the bulk of the absorption, but not the variable features seen at small velocities. We guess that these features are coming from disk outflows because of their moderate correlation with accretion signatures (namely, the He~I NC and He~I AC) as seen in Fig.~\ref{Fig:Corr}. Numerical simulations \citep{romanova_launching_2009, zanni_mhd_2013, pantolmos_magnetic_2020} have shown that the stellar magnetic field lines threading the disk outside the accretion columns are open, further producing transient ejections (i.e., magnetospheric ejections/conical winds) and/or disk winds. These outflows are also suspected when comparing the truncation radius we computed with the size of the region emitting the hydrogen $Br\gamma$ line in S~CrA~N. This region can be spatially constrained to $R_{Br\gamma}=7.7~\pm 2.2~R_*$ thanks to interferometric observations \citep[]{gravity_collaboration_wind_2017, gravity_collaboration_gravity_2023}. The given value is an update from \cite{gravity_collaboration_gravity_2023} with our new values for distance and $R_*$. Since $R_{Br\gamma}$ extends well beyond $R_t$~=~2.1~$\pm$~0.4~$R_*$, disk outflows must account for part of the emission observed in this line.\\
Finally, the Herbig-Haro object HH729 has been attributed to S~CrA~N \citep{peterson_spitzer_2011}. The forbidden lines of $[OI]$ at 6300.2~\AA~ and 6363.8~\AA~ and $[SII]$ at 6730~\AA~ are usually interpreted as tracers of outflows at different scales \citep[see e.g.][]{alexander_dispersal_2014, pascucci_evolution_2020, gangi_penellope_2023}. These lines are present in the spectra of S~CrA~N with a strong blue-shifted peak ($\sim$ -120 km/s) and asymmetry (see Appendix~\ref{app:spectra} for the profiles), strengthening the idea of a large-scale outflow arising from the inner regions.

\section{Conclusions}

Thanks to spectropolarimetric observations in the optical range with ESPaDOnS at CFHT, we have probed the star-disk interaction in the innermost regions of the North component of the young binary system S~CrA. With its confirmed high accretion rate ($10^{-7}$~\msunpyr), this object is an ideal target for studying the magnetospheric accretion scenario in a stronger regime than in CTTS. Our major findings are summarized below: 
\begin{itemize}
    \item When fitting the ESPaDOnS spectra and using CESAM evolutionary models, S~CrA~N appears to have a mass of 0.8~\msun, to be fully convective, and about 1~Myr old. As in previous determinations in the near-infrared range, the veiling we derive around 5500~$\AA$ is strongly variable and high (i.e., ranging from 2 to 11), which suggests that the star experiences a strong accretion regime. Combined with the young age, this could indicate an evolutionary stage between Class I and Class II.
    \item This evolutionary stage is also in line with the large-scale magnetic field we have reconstructed through ZDI. We obtained a total field as strong as 5.4~kG and not strongly axi-symmetric, whose dipolar contribution represents about a third of the total field. Higher poloidal orders being significant ($\sim$ 50\%), the large-scale topology of the magnetic field appears rather complex, when compared with CTTSs. 
    \item Additional developments are needed to paint a complete and coherent view of the magnetic topology of S~CrA~N by including the constraints from the emission lines (such as He~I, the Ca~II IRT or the Fe~II 42 multiplet) and developing more complex models to reproduce the emitting regions of these lines and their properties.
    \item We derive a magnetic truncation radius of $\sim$ 2~$R_*$, and a corotation radius of $\sim$ 6~$R_*$, suggesting that S~CrA~N is in an unstable accretion regime. Looking at the Helium and Hydrogen line profiles and periodicities, we suggest that this accretion occurs in an unstable scenario, along two distinct accretion structures: one main accretion column associated with the accretion shock, and a secondary accretion column with much less density in it as to produce no detectable accretion shock.
    \item The emission lines of the hydrogen Balmer series are highly variable and display multi-component profiles. The observed H$\alpha$ line profiles exhibit clear signatures of an outflow and are in good agreement with simulations of a hot and dense disk wind. This finding is compatible with the size of the $Br\gamma$ emitting region measured with GRAVITY ($\sim$ 8~$R_*$) which is substantially larger than the truncation radius.
\end{itemize}

Our spectropolarimetric campaign in the optical allows us to characterize the star-disk interactions at play in the innermost regions of S~CrA~N and, when combined with near-infrared interferometry, to provide a consistent view of these complex and variable regions. Given the strong and unstable accretion regime, probing the accretion-ejection processes would benefit from a temporal follow-up of these phenomena through simultaneous observations combining photometry, spectroscopy, and interferometry in various spectral ranges.

\begin{acknowledgements}
      \\
      This work is supported by the French National Research Agency in the framework of the "investissements d'avenir" program (ANR-15-IDEX-02).\\
      This work has made use of data from the European Space Agency (ESA) mission {\it Gaia} (\url{https://www.cosmos.esa.int/gaia}), processed by the {\it Gaia} Data Processing and Analysis Consortium (DPAC, \url{https://www.cosmos.esa.int/web/gaia/dpac/consortium}). Funding for the DPAC has been provided by national institutions, in particular the institutions participating in the {\it Gaia} Multilateral Agreement.\\
      This work has made use of the VALD database, operated at Uppsala University, the Institute of Astronomy RAS in Moscow, and the University of Vienna.\\
      Finally, we address our greatest thanks to the referee of this article for their fruitful suggestions and comments.
\end{acknowledgements}

\bibliographystyle{aa}
\bibliography{SCrAN}

\begin{appendix}

\section{Influence of the South component}
\label{app:SCrAS}
In order to address the possible contamination of our observations by S~CrA~S, we simulated the observed Field of View (FoV) of ESPaDOnS. We placed S~CrA~S to a distance of 1.4" from S~CrA~N (astrometry from \citealp{gaia_collaboration_gaia_2022}), and accounted for a circular FoV of 1.6" (from the official ESPaDOnS website\footnote{http://www.ast.obs-mip.fr/projets/espadons/espadons.html}) centered on S~CrA~N. We attributed a normalised flux of 1 for S~CrA~N, hence a flux of 0.63 for S~CrA~S based on a difference in magnitude of 0.5 in the I band \citep{gahm_s_2018}. Finally, we applied a seeing as a 2D Gaussian distribution centered on each star, with its full width at half maximum equal to the seeing of our observations (see Table~\ref{table:log}). This description represents the actual observation and we computed the total flux observed as the sum of the flux in each pixel within the FoV. We also computed the sole contribution of S~CrA~S by modifying the model as such: the FoV is untouched in size and position, but we only take into account the Gaussian model of S~CrA~S (hence not fully included inside the FoV). The contribution of S~CrA~S is then the sum of the flux in each pixel within the FoV. The ratio between these fluxes gives the contribution of S~CrA~S to the total flux observed. This method yields a contribution of $8.96\%$ at worse (seeing of 1350~mas), with a minimum of $0.03\%$ (seeing of 430~mas) and an average of $1.58\%$ (seeing=750 mas). Three observations exceeded a contribution of 5\% : June 26, 27 and 28 with 9.0\%, 9.3\% and 5.3\%, respectively. All the results obtained in this work have been checked excluding these three observations. No significant difference were found, except for a decrease in SNR.

\section{Spectra of S~CrA~N}                                  
\label{app:spectra}
\subsection{Classical emission lines in Classical T Tauri Stars}
\begin{figure*}[t]
    \centering
    \includegraphics[trim=10 0 30 42, clip, width=0.335\linewidth]{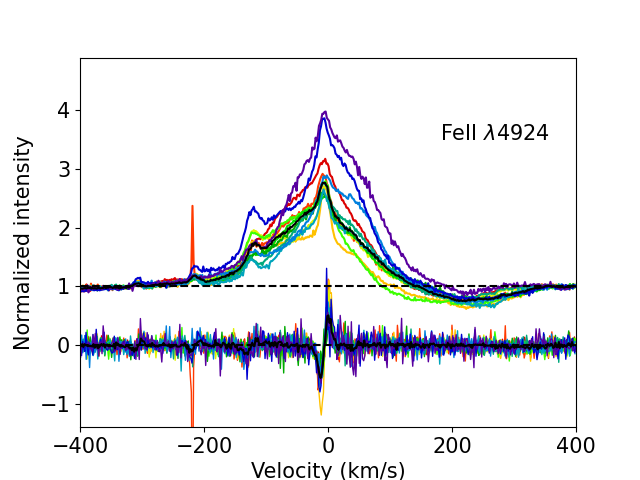}
    \includegraphics[trim=30 0 30 42, clip, width=0.32\linewidth]{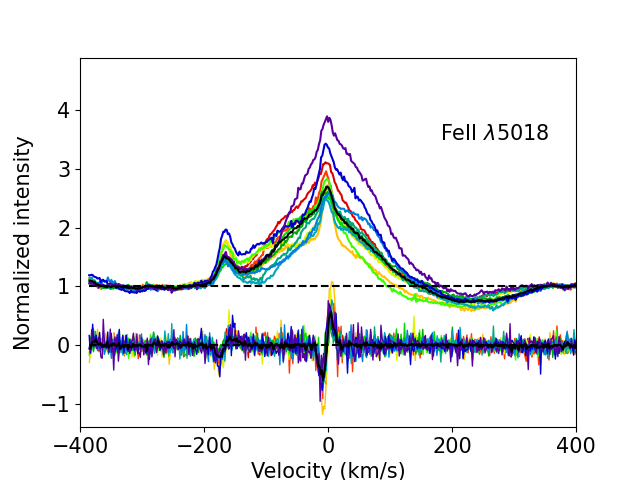}
    \includegraphics[trim=32 0 30 42, clip, width=0.32\linewidth]{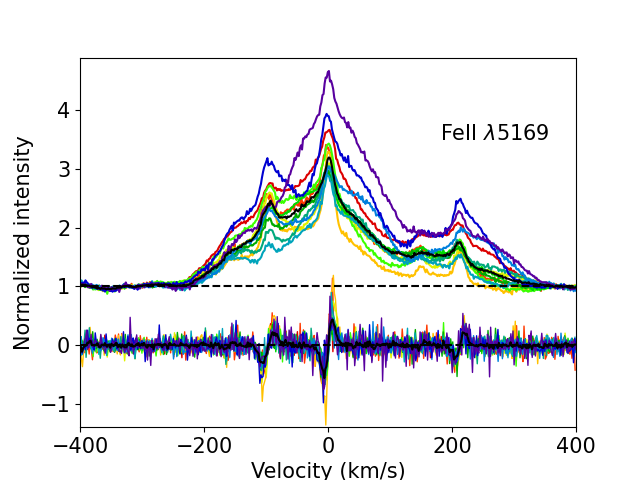}
    \includegraphics[trim=0 0 30 42, clip, width=0.336\linewidth]{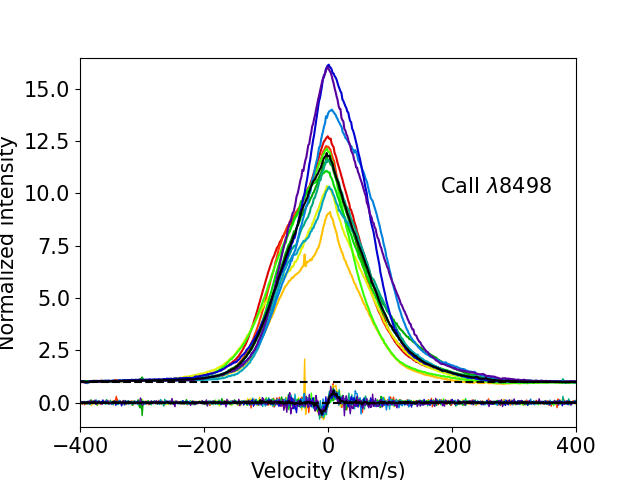}
    \includegraphics[trim=20 0 30 42, clip, width=0.32\linewidth]{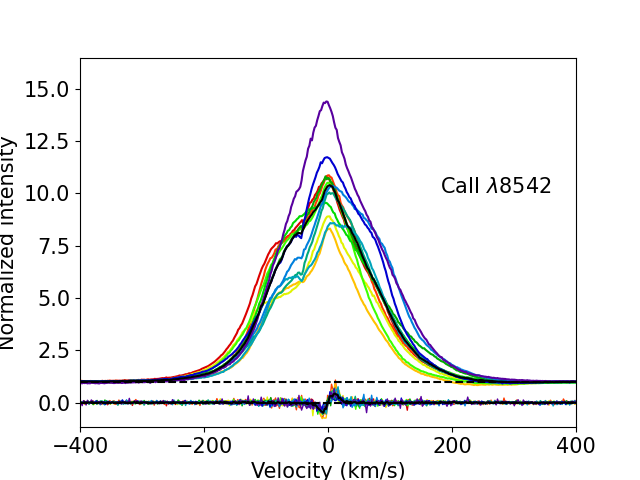}
    \includegraphics[trim=20 0 30 42, clip, width=0.32\linewidth]{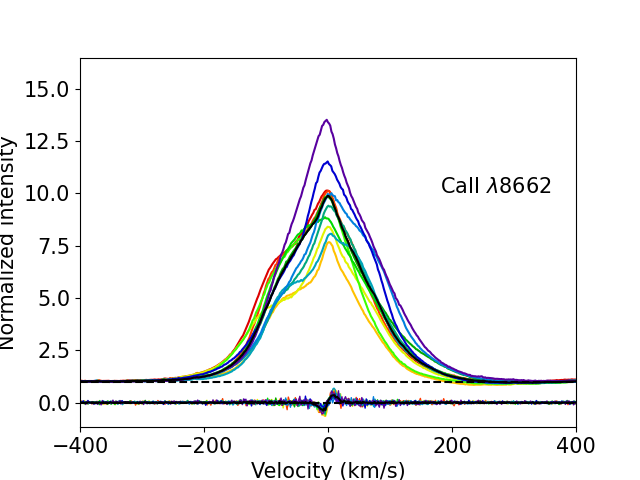}
    \includegraphics[trim=0 0 30 42, clip, width=0.35\linewidth]{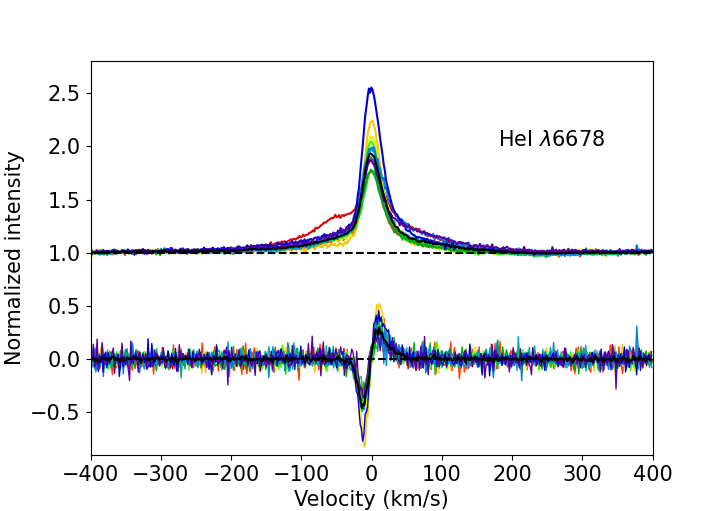}
    \includegraphics[trim=25 0 30 42, clip, width=0.32\linewidth]{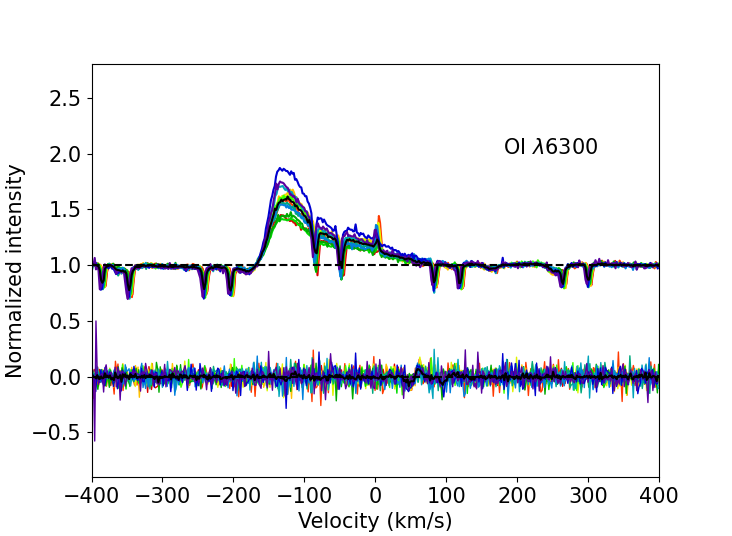}
    \includegraphics[trim=25 0 30 42, clip, width=0.32\linewidth]{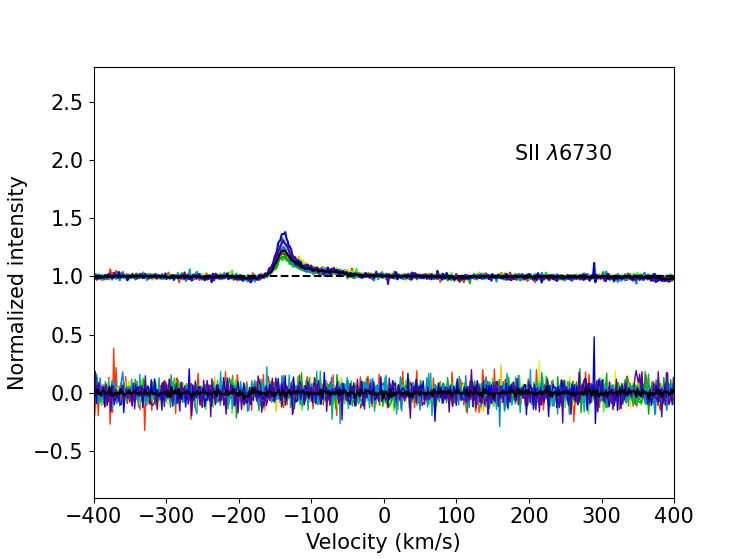}
    
    \caption{Emission lines in S~CrA~N. All $V$ profiles have been magnified by a factor 5, for clarity. Rainbow shading has the same meaning as in Figure~\ref{Fig:spectra_zoom}, solid black line is the median profile.}
    \label{Fig:emission_lines}
\end{figure*}
Some lines classically tracing the accretion-ejection processes in CTTS are displayed in Figure~\ref{Fig:emission_lines}. We provide a short comment on the aspect and behaviour for each of these lines.\\
\subsubsection*{\bf Iron triplet}
This triplet thought to originate from the chromospheric activity of the star is presented in the top row of Fig.~\ref{Fig:emission_lines}. They show great variability and strong intensity. They all exhibit two common features : A broad and a narrow emission, both centered near the radial velocity of the star. The lines at $\lambda4924$ and $\lambda5018$ also have an additional red-shifted absorption at high velocities, likely to be interpreted just like the ones observed in He~I~$\lambda5876$, $H\gamma$ and $H\delta$. Their Stokes $V$ profiles show anti-symmetric signatures associated to the narrow emissions which are exceptionally strong and symmetric. Combined with the simple decomposition of their Stokes $I$ counterparts, this makes them good candidates for a future tomographic reconstruction taking emission lines as constraints.
\subsubsection*{\bf Calcium infrared triplet (IRT)}
Commonly used to study the activity of CTTS, this triplet is presented in the middle row of Fig.~\ref{Fig:emission_lines}. Its intensities are comparable to those of H$\alpha$, which put these lines amongst the most dominant ones in the spectrum of S~CrA~N. Their shape is complex and no simple decomposition (based on the sum of up to 4 Gaussians) was successful in reproducing them. It should be noted that the Stokes V signatures are strong and define a narrow component emitted near the surface of the star, unlike the rest of the lines.
\subsubsection*{\bf He~I $\boldsymbol{\lambda6678}$}
Likely originating from the same processes as He~I $\lambda5876$. Its different excitation potential is used to compute line ratios and infer the electron density and temperature encountered in the region producing these lines. No attempt to derive these parameters was tried in this work. The decomposition of this line appears similar to that of He~I $\lambda5876$, albeit the noticeable absence of the red-shifted AC.
\subsubsection*{\bf Oxygen}
This forbidden line of oxygen is usually threefold \citep[see e.g.][]{pascucci_evolution_2020} : A narrow low velocity component (NLVC) traces a wind arising in the outermost part of the inner disk ($\sim$ 1 au). A broad low velocity component (BLVC) traces a wind arising in the innermost part of the inner disk ($\leq$ 0.1 au). A high velocity blue-shifted component (HVC) traces a jet. The HVC is observed peaking around -120 km/s and confirmed by a Herbig Haro object (HH729) associated to S~CrA~N. Regarding the low velocity components, S~CrA~N seems to display a BLVC with no NLVC, which is unexpected since disk evaporation models predict an evolution going from BLVC + NLVC (when the inner wind is dense enough to shield the outer disk from high energy stellar photons) to NLVC as the inner wind's density decreases and the disk starts evaporating \citep{alexander_dispersal_2014}. Since the disk is not expected to evaporate before the late class II stage, another model should be invoked to explain the profiles observed in S~CrA~N (class I - class II transition stage). No Stokes V signal is detected, as expected from a line formed away from the star, in a region where the magnetic field is weak.
\subsubsection*{\bf Sulfur}
Just like for the forbidden oxygen line, this line is tracing a jet, with a peak at similar velocities as the HVC of OI : $\sim$ -120 km/s.

\subsection{Variability Of the Balmer series}
\begin{figure}
    \centering
    \includegraphics[width=0.49\linewidth]{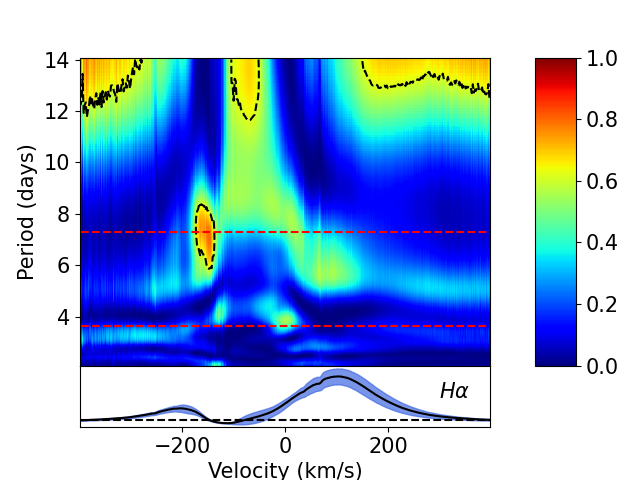}
    \includegraphics[width=0.49\linewidth]{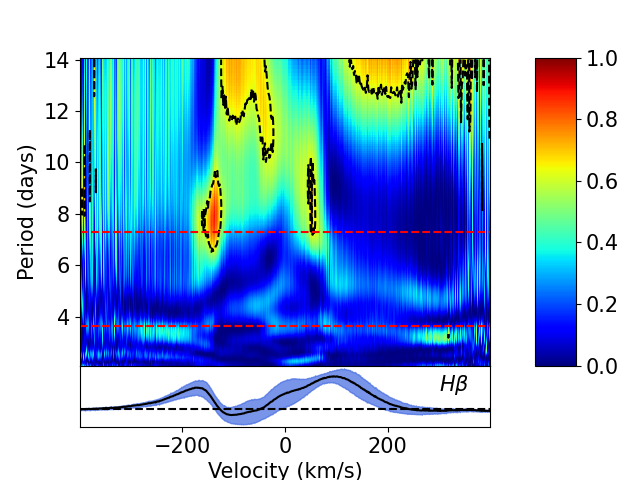}
    \includegraphics[width=0.49\linewidth]{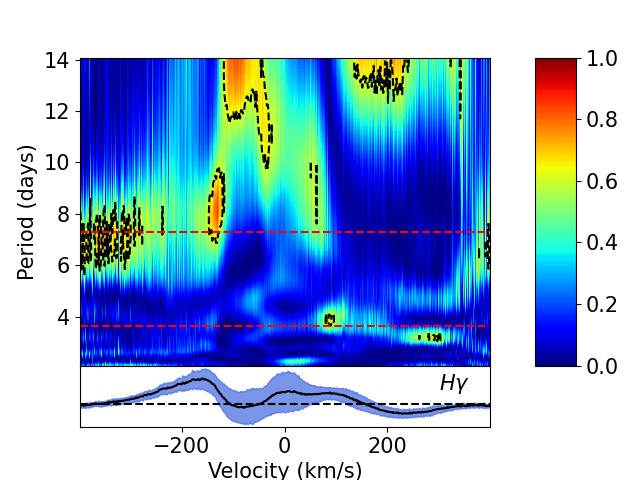}
    \includegraphics[width=0.49\linewidth]{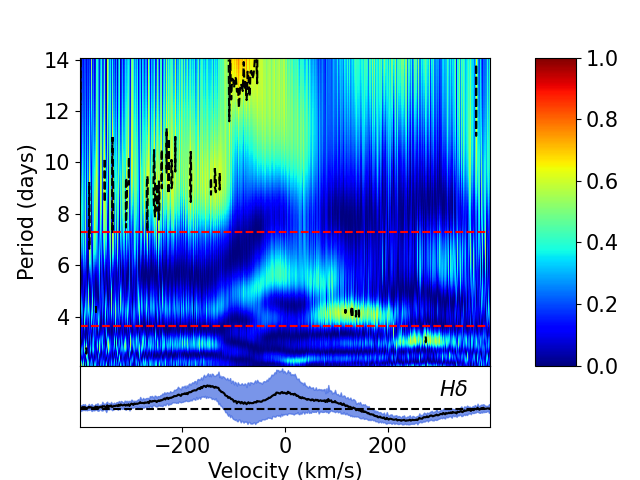}
    \caption{2D periodograms of the lower Balmer series. Red dashed lines show periods of 7.3 days ($P_*$) and 3.65 days ($P_*/2$). Black dashed contour show a 3\% FAP level.}
    \label{fig:2DP_Balmer}
\end{figure}
The Balmer series are composite lines, as discussed in this work. Their 2D periodograms are shown in Fig.~\ref{fig:2DP_Balmer}. Multiple signals can be seen, but as discussed in the core of this work, the higher periods will not be commented, because of the white noise assumption made in the computation of the periodograms, and because of the total span of observation, which does not allow to observe these periodicities for more than a full cycle. The blue side of the broad emission (near -180 km/s) has a periodicty of the order of stellar rotation period ($7.3~\pm~0.2$~days) in all the lines but $H\delta$. The red side of this broad absorption (near +100 km/s) has no clear periodicity for $H\alpha$ and $H\beta$, while a somewhat significant signal near 4 days can be seen for $H\gamma$ and $H\delta$. This periodicity is found nowhere in any other feature we studied in this work, and its origin remains unknown. The red-shifted absorption seen in $H\beta$, $H\gamma$ and $H\delta$, which has been extensively discussed in previous sections is found to be periodic with a period similar to the one in the red-shifted absorption met in the He~I $\lambda5876$ line. That is, slightly lower than half the rotation period (3.2 days instead of 3.65 days)
\subsection{LSD profiles dependence on the lines depth}

\begin{figure}[ht]
    \centering
    \includegraphics[trim=18 8 39 40,clip,width=\linewidth]{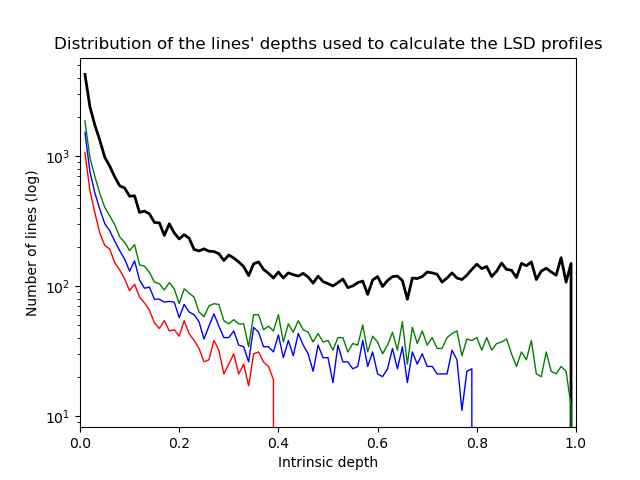}
    \includegraphics[trim=9 48 28 71,clip,width=0.49\linewidth]{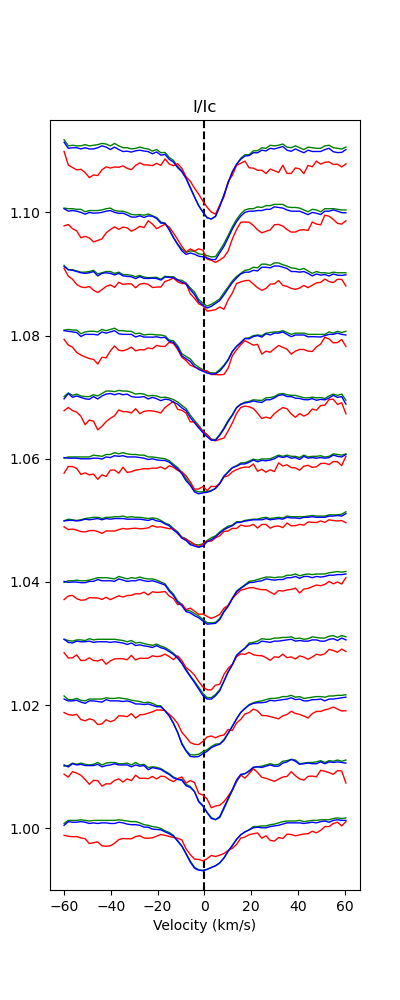}
    \includegraphics[trim=9 48 28 71,clip,width=0.49\linewidth]{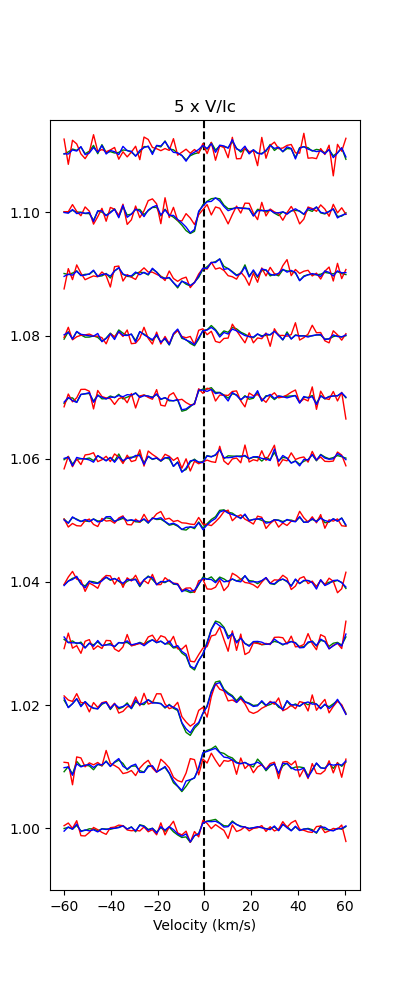}
    \caption{Illustration of the lines depth test. Top : Number of lines included in each mask, as a function of their intrinsic depth. Bottom-left : LSD $I$ profiles associated to each mask. Bottom-right : LSD $V$ profiles (magnified by a factor 5, for clarity) associated to each mask. Black curve stands for the raw synthetic spectrum. Green curve corresponds to the mask used in this study. Blue curve corresponds to a depth limit of 0.8. Red curve corresponds to a depth limit of 0.4.}
    \label{Fig:MultiD}
\end{figure}

In order to test the potential influence of the depth of the lines taken into account during the LSD procedure, we performed a test where we excluded the deepest lines, which should be the most affected by veiling \citep{rei_line-dependent_2018}, and the ones located within their $\pm$ 15 km/s vicinity, which could be affected by them. We applied different depths thresholds to obtain a mask, and computed the resulting LSD $I$ and $V$ profiles. We illustrate this with 3 different masks on Fig.\ref{Fig:MultiD}. The first mask is the one we cleansed naturally, i.e. without any depth threshold (green curves). It includes 10 263 lines out of the 27,502 coming from the raw synthetic spectrum. Then, with a threshold of 0.8, only 7,301 lines were left in the mask (blue curve). And finally, with a threshold of 0.4, 4,279 lines remained (red curve). At this stage, the profiles' SNR becomes so low that any further cut would significantly impact the  the LSD signal. It is visible from Fig.~\ref{Fig:MultiD} that no matter the depth limit applied to a mask, the qualitative shape of the LSD profiles remain constant. We can then exclude the continuum veiling as the source of the quick and intense variations observed in our LSD profiles, and favour an origin in the chromospheric emission lines.

\section{Distance estimate of the S~CrA system}
\label{app:dist}

\begin{figure}[ht]
    \centering
    \includegraphics[width=\linewidth]{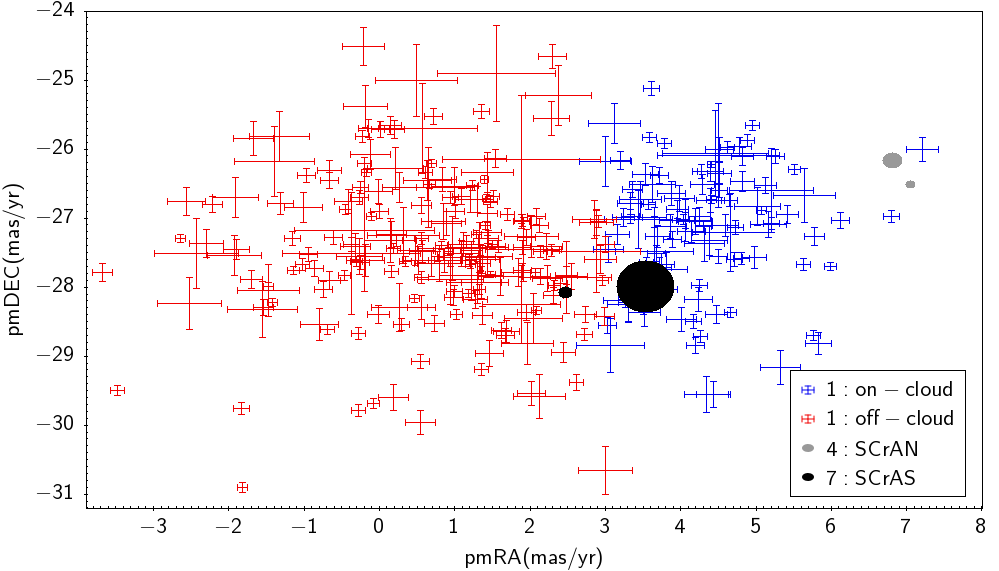}
    \caption{Proper motions for the two subgroups of the Corona Australis region in Gaia DR2, S~CrA~N and S~CrA~S in Gaia DR2 and Gaia DR3. Adapted from \textbf{\cite{galli_corona-australis_2020}}.}
    \label{Fig:distSCrAN}
\end{figure}

The distance found for S~CrA~N in the most recent literature comes from \cite{prato_astrophysics_2003} with $130\pm20$~pc. This value is taken from an even earlier estimate of the distance of the whole R~CrA~T region by \cite{marraco_distance_1981}, while the associated uncertainty is simply assumed \textit{a priori}, and does not derive from the observations. The latest Gaia-DR3 data \citep{gaia_collaboration_gaia_2022} provides a distance of $160 \pm 2$~pc for S~CrA~N from its parallax measurements\footnote{S~CrA~N Gaia ID : 6731210253964165632}, while S~CrA~S is found to be at $147 \pm 2$~pc with the same data release\footnote{S~CrA~S Gaia ID : 6731210258265107584}, which is inconsistent with the assumption of S~CrA being a close binary. This discrepancy alone is not enough to question the binarity of the S~CrA system though. The nebulosity in its surroundings, coupled with its binary nature, can explain such a discrepancy in the Gaia parallax measurements. \cite{galli_corona-australis_2020} used the Gaia-DR2 data on a sample of 313 stars associated with the Corona Australis region to draw a line between the so-called "on-cloud" and "off-cloud" sub-regions. In their study, this distinction is based on a Bayesian distance estimate and tangential velocities measurements. Using the Gaia-DR3 data, we placed S~CrA~S and S~CrA~N in the diagram in Figure~\ref{Fig:distSCrAN} (proper motion in right ascension against proper motion in declination for both Gaia DR2 and DR3 data) adapted from \cite{galli_corona-australis_2020} and non-ambiguously assigned S~CrA~N to the "on-cloud" region with a proper motion of $7.05 \pm 0.06$ mas/yr in right ascension and $-26.51 \pm 0.06$ mas/yr in declination. We thus adopted the distance of the "on-cloud" region, that is $152.4 \pm 0.4$ pc for S~CrA~N.

\end{appendix}

\end{document}